\documentclass[lettersize,journal]{IEEEtran}
\usepackage{amsmath,amsfonts}
\usepackage{algorithmic}
\usepackage{algorithm}
\usepackage{array}
\usepackage[caption=false,font=normalsize,labelfont=sf,textfont=sf]{subfig}
\usepackage{textcomp}
\usepackage{stfloats}
\usepackage{url}
\usepackage{verbatim}
\usepackage{graphicx}
\usepackage{epstopdf}
\usepackage{cite}
\usepackage{hyperref}
\usepackage{threeparttable}
\def\BibTeX{{\rm B\kern-.05em{\sc i\kern-.025em b}\kern-.08em
    T\kern-.1667em\lower.7ex\hbox{E}\kern-.125emX}}
\usepackage{balance}

\begin{document}

\title{A Lightweight Authentication Protocol against Modeling Attacks based on a Novel LFSR-APUF}
\author{Yao Wang,~\IEEEmembership{Member,~IEEE}, Xue Mei, Zhengtai Chang, Wenbing Fan, Benqing Guo, and Zhi Quan 
\thanks{This work was supported by the National High Technology Research and Development Program of China under grant No. SQ2020YFF0404465. \emph{(Corresponding author: Yao Wang and Zhi Quan)} \par
Yao Wang, Xue Mei, Zhengtai Chang, Wenbing Fan and Zhi Quan are with the School of Electrical and Information Engineering, Zhengzhou University, Zhengzhou 450001, China (e-mail: ieyaowang@zzu.edu.cn; mx690774079@163.com; changzhengtai@163.com; iewbfan@zzu.edu.cn; iezquan@zzu.edu.cn).\par
Benqing Guo is with the College of Communication Engineering, Chengdu University of Information Technology, Chengdu, China(e-mail: rficgbq@gmail.com).}}

\markboth{IEEE INTERNET OF THINGS JOURNAL,~Vol.~18, No.~9, September~2020}%
{How to Use the IEEEtran \LaTeX \ Templates}

\maketitle

\begin{abstract}
Simple authentication protocols based on conventional physical unclonable function (PUF) are vulnerable to modeling attacks and other security threats. This paper proposes an arbiter PUF based on a linear feedback shift register ({LFSR-APUF}).\ Different from the previously reported linear feedback shift register for challenge extension, the proposed scheme feeds the external random challenges into the LFSR module to obfuscate the linear mapping relationship between the challenge and response.\ It can prevent attackers from obtaining valid challenge-response pairs (CRPs),\; increasing its resistance to modeling attacks significantly. A 64-stage {LFSR-APUF} has been implemented on a field programmable gate array (FPGA) board. The experimental results reveal that the proposed design can effectively resist various modeling attacks such as logistic regression (LR), evolutionary strategy (ES), Artificial Neuro Network (ANN), and support vector machine (SVM) with a prediction rate of 51.79\% and a slight effect on the randomness, reliability, and uniqueness. Further, a lightweight authentication protocol is established based on the proposed {LFSR-APUF}.\ The protocol incorporates a low-overhead, ultra-lightweight,  novel private bit conversion \emph{Cover} function that is uniquely bound to each device in the authentication network.\ A dynamic and time-variant obfuscation scheme in combination with the proposed {LFSR-APUF} is implemented in the protocol. The proposed authentication protocol not only resists spoofing attacks, physical attacks, and modeling attacks effectively, but also ensures the security of the entire authentication network by transferring important information in encrypted form from the server to the database even when the attacker completely controls the server. 
\end{abstract}


\section{Introduction}\label{sec1}
\IEEEPARstart{T}{he} Internet of Things (IoT) connects and integrates the physical world and digital space based on the existing and evolving information and communication technologies, 
and it is usually considered a worldwide network of interconnected objects uniquely addressable based on standard communication protocols\cite{IOT2014}.
However, as the scale of IoT continues to expand and the number and types of access devices gradually increase, information security in IoT devices is becoming a major concern for users. 
This issue can be addressed by authentication protocols based on various encryption algorithms such as AES and RSA \cite{bib23}. The secret key, which is critical for the encryption algorithms and should remain strictly confidential, is commonly stored in non-volatile memory (NVM). However, some IoT devices are highly resource-constrained with low cost, small size, and ultra-low-power consumption, making it difficult to implement complex encryption algorithms. Furthermore, the NVM storing the secret key is vulnerable to various physical attacks, including side-channel, semi-invasive, and invasive attacks \cite{bib24},\cite{IOT2019}.\par
Physical unclonable functions (PUFs) are promising hardware security primitives for resource-constrained systems \cite{bib1}. 
PUFs utilize the inevitable random physical variations during the chip manufacturing process to extract a unique bitstream. 
Specifically, PUF employs the uncontrollable manufacturing deviations 
at the deep submicron level to produce unique challenge-response pairs (CRPs) as fingerprints for each instance.\, The ability to control the precise generation time of the secret bitstring and the sensitivity of the PUF entropy source to invasive probing attacks (which would cause changes in the electrical characteristics of the PUF circuits and make the PUF invalid) are additional attributes that make them attractive for authentication in embedded hardware.\,
PUFs can be divided into two categories: strong PUFs (SPUFs) and weak PUFs (WPUFs) \cite{bib26}. 
SPUFs, represented by arbiter PUF (APUF), have more CRPs and are usually employed for authentication purposes. \par

\IEEEpubidadjcol

Although SPUF-based protocols provide an extremely effective and lightweight solution 
for the authentication of IoT devices, 
they are not completely reliable and are vulnerable to some serious security attacks such as modeling attacks. 
\, Firstly, the attacker collects a part of CRPs of the SPUF, 
then employs machine learning algorithms to establish a mathematical model of the SPUF, 
and finally predicts unknown CRPs with high accuracy. \, In this way, the attackers can "clone" a SPUF. 
Back in 2010, R\"{u}hrmair et al. used computational algorithms to simulate some PUF behaviors for the CRP dataset, 
and then they used logistic regression (LR) and evolution strategies (ES) 
to successfully break standard APUFs and ring oscillator PUFs as well as XOR APUFs, feed-forward APUFs, and lightweight Secure PUFs \cite{bib25}.\,
Especially, LR could break 64-stage APUF with 99.9\% prediction accuracy based on over 18,000 CRPs.\, 
Later in 2013, R\"{u}hrmair et al. performed modeling attacks on standard APUF and XOR APUF 
by obtaining silicon CRPs data from field programmable gate arrays (FPGAs) 
and application-specific integrated circuits (ASICs) \cite{bib2}. 
The experimental results from silicon CRPs were very close to 
those from simulated CRPs, which not only validated the accuracy 
of the inference drawn from earlier versions of the work 
but also revealed the vulnerability of SPUF to machine learning attacks (MLAs). \par

To protect SPUF-based authentication protocols from possible modeling attacks, 
some modified structures have been reported. These structures can be roughly divided into two categories: 
enhancing the PUF structure’s nonlinearity or obfuscating the CRPs. 
Each delay chain of conventional APUF is the linear cumulative total of its delay element, 
causing vulnerability to modeling attacks.\ To break the linearity of the conventional APUF and enhance its nonlinearity, 
particular nonlinear components are introduced into APUF \cite{bib3},\cite{bib4},\cite{bib5}. However, variations in the PUF’s internal elements are generally unsuitable for FPGA implementation. 
CRPs obfuscation can also improve the MLA resistance ability of PUF. This technique does not require the underlying structural variations of PUF. 
It only introduces specific mapping circuits at the challenge input or response output ports of the PUF core 
to obfuscate the real mapping relationship between the challenges and responses, which prevents the attackers from obtaining effective CRPs \cite{bib6},\cite{bib7},\cite{bib8},\cite{bib9}. However, obfuscation modules, especially encryption function modules, 
usually have significant hardware overhead. 
Besides, the obfuscation technique essentially employs a fixed mapping 
that cannot effectively resist advanced MLAs \cite{bib11},\cite{bib12}. \par

In addition to the SPUF-based authentication protocol’s vulnerability to modeling attacks, 
most of the above protocols \cite{bib6},\cite{bib7},\cite{bib8},\cite{bib9}
assume that the server is secure during the authentication phase, i.e. , 
the server cannot be compromised by any attacker. 
This is because the server stores CRPs regarding SPUF or the software model of SPUF. 
Once the server is broken in the authentication network, 
the attacker can obtain information related to the SPUF and other encryption functions, 
which can easily destroy the entire authentication protocol.\par

In this study, an APUF based on a linear feedback shift register (LFSR), called {LFSR-APUF}, 
is proposed to increase the resistance to modeling attacks.\  Furthermore, a new lightweight authentication protocol against modeling attacks is established based on the {LFSR-APUF}.\  The significant contributions of the work are summarized as follows:\par

1) An obfuscation scheme is proposed, which can be utilized with existing SPUFs to improve the modeling attack resistance. FPGA experiments indicate that the proposed {LFSR-APUF} significantly decreases the prediction rate of modeling attacks. The prediction rate of logistic regression (LR), evolutionary strategy (ES), Artificial Neuro Network (ANN), and support vector machine (SVM) can be reduced to nearly 50 \% by setting relevant parameters.\par

2) A 64-stage {LFSR-APUF} is implemented on the {Xilinx Artix-7 FPGA} board. An unbiased APUF hard macro design methodology is described. This methodology not only ensures the symmetry of the two delay paths of the underlying APUF but also facilitates the modification of obfuscation logic such as LFSR without impacting the delay of the underlying APUF. Furthermore, the randomness, uniqueness, reliability, and authentication capacity of the {LFSR-APUF} circuit are evaluated.\par

3) A new lightweight authentication protocol is proposed, which makes full use of the special characteristic of {LFSR-APUF} to implement the dynamic obfuscation scheme and further improve the protocol-level modeling attack resistance. The proposed protocol model includes several devices, a server, and a database, any of which can be attacked by attackers. To protect the transmission of CRP information over communication links (i.e.\,, between devices and server as well as between server and database), a new ultra-lightweight bitstream encryption function \emph{Cover} is established.\par

The remainder of this paper is arranged as follows.
Section \ref{sec2} describes the modeling attacks and the corresponding security strategy of APUF. 
Section \ref{sec3} introduces the proposed {LFSR-APUF}.
The experimental results, including various modeling attacks, randomness, reliability, and uniqueness 
are presented in Section \ref{sec4}.
Section \ref{sec5} presents the proposed lightweight authentication protocol based on {LFSR-APUF}.\,  
The security of this authentication protocol is analyzed in Section \ref{sec6}.\,
Finally, the study is concluded in Section \ref{sec7}.

\section{Related Work}\label{sec2}

\begin{figure}[!t]
\centering
\includegraphics[width=\columnwidth]{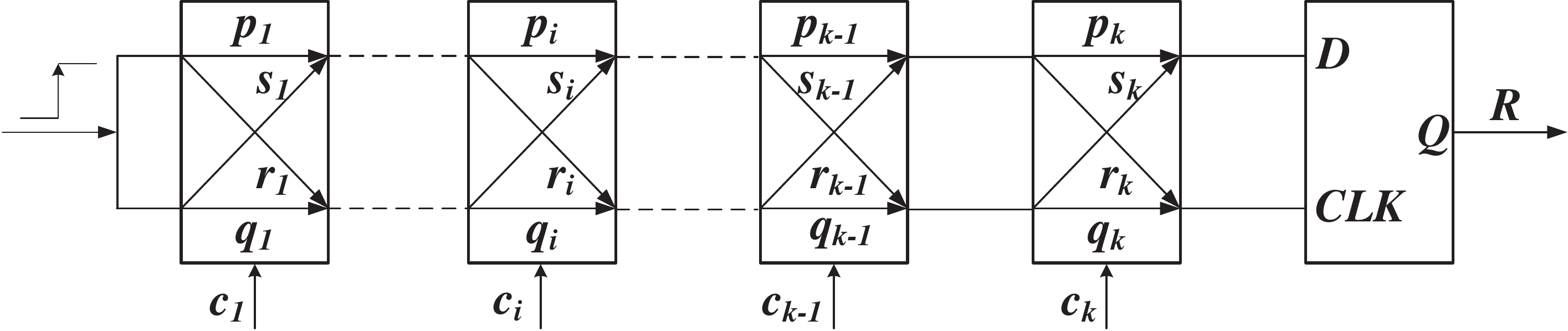}
\caption{The structure diagram of k-stage APUF}\label{fig1}
\end{figure}

As shown in Fig. \ref{fig1}, conventional APUF \cite{bib1} consists of $k$-stage
cascaded switch blocks, and all of them are entirely
symmetric to ensure that the two paths have similar nominal
delays. The challenge $c_i$ of APUF acts as the selected signal of the switch block's path
to form two different delay chains. If $c_i$ is 0, the
upper and lower pulses pass through the switch block in parallel;
if $c_i$ is 1, the upper and lower pulses cross through the switch
block. The D flip-flop at the terminal of two delay chains
generates a response according to the arrival time difference 
between the top and bottom pulses.\, If the top pulse arrives at the
D port of D flip-flop earlier, $R$ is equal to 1; otherwise, $R$ is 0.\par

Although a conventional $k$-stage APUF can generate $2^k$ CRPs,
different challenges have the same $k$-stage switch blocks.
Therefore, the conventional APUF may be considered a linear
model, causing a strong correlation between different
CRPs. As shown in Fig.\ref{fig1}, the delay of the four signal paths of the $i-th$
switch block are denoted by $p_i$, $q_i$, $s_i$, and $r_i$, and the final $R$ can
be formulated as \cite{bib2}:

\begin{equation}
R=sign(\Delta)=sign(\overrightarrow{\Omega^T}\overrightarrow{\Phi})\label{eq11}
\end{equation}
Where $sign (x)$ is the sign function,when $x$ \textless 0, $sign (x)$=0, otherwise $sign (x)$=1.

\begin{equation}
\overrightarrow{\Omega}=(\omega^1,\omega^2,\dots,\omega^{n+1}) \label{eq22}
\end{equation}

Where,
\begin{equation}
\omega^1= \frac{\delta^0_1-\delta^1_1}{2},w^i= \frac{\delta^0_{i-1}+\delta^1_{i-1}+\delta^0_i-\delta^1_i}{2} \label{eq33}
\end{equation}

\begin{equation}
 \delta^0_i=p_i-q_i, \delta^1_i=s_i-r_i \label{eq44}
\end{equation}

\begin{equation}
\overrightarrow{\Phi}=(\Phi^1,\Phi^2,\dots ,\Phi^k,1)^T \label{eq55}
\end{equation}

\begin{equation}
\Phi^l=\prod^k_{i=1}(1-2c_i) \label{eq66}
\end{equation}

$\Omega$ provides a compact model for the process variation of PUF. $\Phi$ is the result of a reversible challenge transformation. If $\Delta=\overrightarrow{\Omega^T}\overrightarrow{\Phi}=0$,
it can be abstracted into a mathematical k-dimensional hyperplane. Any challenge falling on one side of the hyperplane generates the response “0”; otherwise, the response is “1”. The attacker can fit this hyperplane by learning some known CRPs through machine learning, and this hyperplane can help him/her predict the corresponding response bit for unknown challenges.\par

The nonlinearity of the PUF structure can be utilized to improve the modeling attack resistance. Some structural nonlinearization circuits such as current-mirror PUF and voltage transmission PUF \cite{bib3},\cite{bib4},\cite{bib5} adopt the nonlinear characteristics of current mirrors and circuit transfer function to increase the modeling attack resistance. However, the underlying essential elements of the PUF circuit are required to be modified, which cannot be achieved on FPGAs, and their control is difficult.\, In feed-forward APUF, several challenges are generated by the intermediate stages' arbiters rather than those connected with external signals. These PUFs introduce structural nonlinearization to resist the traditional MLAs such as LR. Unfortunately, their stability is weak due to the nonlinearization of PUF elements. Besides, these nonlinear PUF structures are vulnerable to ES-based heuristic MLAs \cite{bib29}.\par

CRP obfuscation adds pre/post-processing circuits at challenge input or response output ports to hide the natural mapping relations of CRP. Therefore, the attackers can only collect obfuscated “CRPs”, while the real ones cannot be accessed. Various anti-MLA SPUFs with CRP obfuscation have been proposed in the literature.\, Controlled PUF (CPUF) \cite{bib28} employs random hash to encapsulate the PUF core, thereby cutting off the direct relationship between the response of PUF core and CPUF output. To enhance the reliability, an error correction code (ECC) module is used, resulting in the cost of area and power overhead. In particular, the leakage of helper data from the ECC module makes PUF vulnerable to reliability-based modeling attacks. PUF finite state machine (PUF-FSM) \cite{bib6} removes the hash logic and replaces the ECC module with FSM. The response generated by the PUF is fed to the FSM, which then iterates through a given set of states under the control of the response bitstream until it reaches the final correct state. If an attacker applies the false challenges to the PUF, the generated responses prevent the FSM from reaching the final state, and therefore the legitimate hash output cannot be generated. However, PUF response in PUF-FSM passes through the hash function without ECC, which may lead to an avalanche of the hash output due to the flipping of individual unstable bits during the
authentication phase. In addition, the hash logic at the response port still imposes significant overhead, and PUF-FSM obfuscation is compromised by the covariance matrix adaptation evolutionary strategies (CMA-ES) \cite{bib11}. Lockdown 
scheme\cite{bib30} limits the number of CRPs available to an attacker by setting an upper limit on the number of CRPs. Only the trusted server can access the new CRPs. However, this approach limits the number of CRPs used for authentication and does not allow any repeated challenge bitstreams. Therefore, this protocol is not suitable for applications that require frequent data exchange and fast authentication. A random number generator (RNG) is introduced before a SPUF to randomize the challenge bits, so the MLA resistance of PUF with the randomized challenge (RPUF) \cite{bib7} is enhanced. Notably, the above obfuscation schemes can be applied to most existing PUFs to enhance their anti-MLA ability.\, Unfortunately, both PUF-FSM and RPUF have been broken \cite{bib11},\cite{bib12}. Recently, a random set-based obfuscation for strong PUF (RSO-PUF) has been proposed in \cite{bib8}. The structure uses the stable response from the device PUF as the obfuscation key. A specific amount of stable challenges are stored in an NVM inside the device. During the authentication phase, these challenges, stored as obfuscation keys in the register within the device, are used to generate the responses. The true RNG (TRNG) randomly selects two keys from the register. The first key is XORed with the original challenge, and the second key is XORed with the original response. The obfuscated response obtained by the attacker is the XOR value of the original response and the second key. The modeling attack resistance heavily depends on NVM, which increases the hardware overhead significantly and makes the memory vulnerable to be read by any attacker. \, In general, all the above PUF-based authentication protocols rely on the assumption that the server is unbreakable, which is difficult to guarantee in real IoT applications. Once the server is broken, all the information related to the PUF in the embedded device is exposed, posing a threat to the entire network. PUF identity-preserving protocol (PUF-IPA) \cite{bib31} assumes that the server can be broken by an attacker, so the server does not store the CRP dataset or the soft model of the PUF that is embedded in the device during the authentication phase. However, PUF-IPA achieves server security at the expense of device security. The NVM in the device exposes the device to the threat of memory reading. Once the NVM in the device is reset by the attacker, the authentication protocol can be easily broken by replay attack. In addition, the NVM, hash logic, and other heavyweight circuits in the device significantly increase the hardware overhead. Modeling attack resistant deception technique \cite{bib35} uses an active deception protocol to prevent machine learning (ML) attacks. However, using two Strong PUFS increase the 
hardware overhead of the device. \par

\section{Proposed LFSR-APUF}\label{sec3}

The conventional time-invariant obfuscation scheme converts
the external challenge into the obfuscated challenge with some
static mapping, while the proposed time-variable
obfuscation scheme changes the mapping according to
the external challenge in real-time.

\subsection{Design of Proposed LFSR-APUF}\label{subsec31}

\begin{figure}[!t]
\centering
\includegraphics[width=\columnwidth]{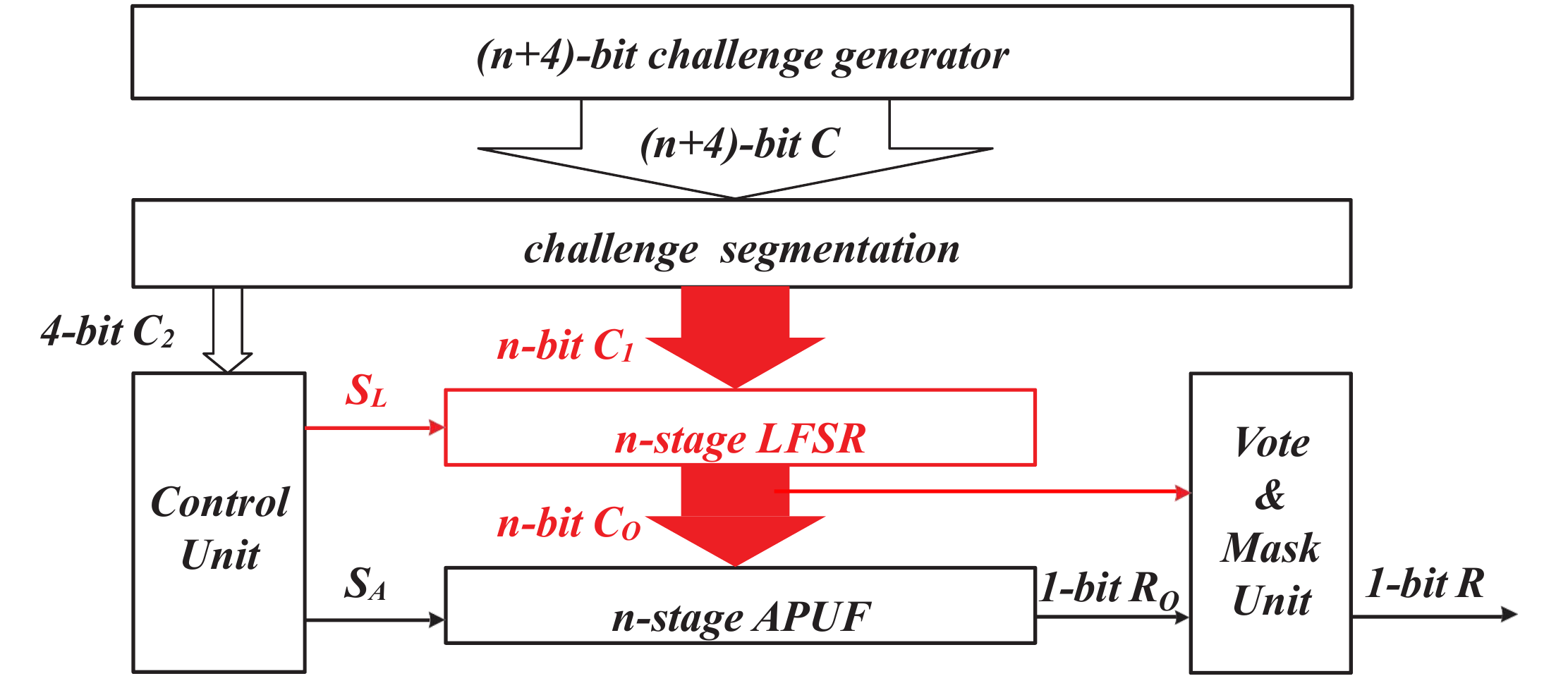}
\caption{The schematic diagram of proposed the LFSR-APUF}\label{fig2}
\end{figure}

As shown in Fig. \ref{fig2}, the proposed LFSR-APUF can be
divided into the following six parts:\par

\textbf{Challenge Generator}: Mainly generates random external challenges from outside during implementation on FPGA board.\par

\textbf{chanllenge segmentation} : Divide $(n+4)$-bit 
 $C$ into two parts. Part of $n$-bit $C_1$ is passed to {LFSR}. Another part of the 4-bit $C_2$ is applied to Control\; Unit.\par

\textbf{APUF Unit} : A conventional APUF circuit with n-stage symmetrical switch blocks.\par

\textbf{LFSR Unit} : Mainly obfuscates the external challenges by several numbers of shifts.\par

\textbf{Control Unit} : Mainly controls the LFSR shift according to the external sectional challenge, generates an obfuscated challenge, and controls the APUF to generate a response bit.\par

\textbf{Vote \&\ Mask Unit} : Voting mechanism and soft dark mask \cite{bib34} are used on APUF.\, Mainly screen out unstable {CRP}s to improve reliability.\par

According to the above structure, the specific working
process of the proposed LFSR-APUF is summarized as
follows:\par

The $Challenge\;Generator$ generates $(n+4)$-bit challenge $C$,
where $n$-bit $C_1$ and the remaining 4-bit challenge $C_2$ are
considered the inputs of the $n$-stage $LFSR\;Unit$ and the
$Control\;Unit$, respectively. Note that $C_1$ is not all-zeros or all-ones, otherwise the LFSR will be stuck.\par

The $n$-stage $LFSR\;Unit $ receives the $n$-bit $C_1$ generated by the
$Challenge\;Generator$ and considers it as the initial value. 
Then, it waits for the shift signal $S_L$ sent by the $Control\;Unit$.\par

The $Control\;Unit$ receives the remaining $4$-bit $C_2$ and
determines the number of shifts of the LFSR. For example, when $C_2$ is equal to 0110, the $LFSR$ should
be shifted six times, and its final output is employed as the $n$-bit obfuscated challenge $C_O$. Finally, when the
$Control \; Unit$ sends the shift signal $S_L$ a specific number of
times, it sends the pulse signal $S_A$, namely the rising edge
transition signal, to the $n$-stage $APUF$.\par

The $n$-stage APUF receives $n$-bit obfuscated challenge $C_o$
generated by $LFSR$ and the $Control\;Unit’s$ pulse signal $S_A$. 
As shown in Fig. \ref{fig1}, when the pulse signal is
transmitted through two delay chains from the start to the
final D flip-flop, the $APUF$ generates a one-bit response $R_o$ once. $Vote \&\ Mask\; Unit$  screens out unreliable CRPS and outputs reliable one bit response $R$.

\subsection{Segmentation Strategy of the Proposed LFSR-APUF}\label{subsec32}

\begin{figure}[!t]
\centering
\includegraphics[width=2.5in]{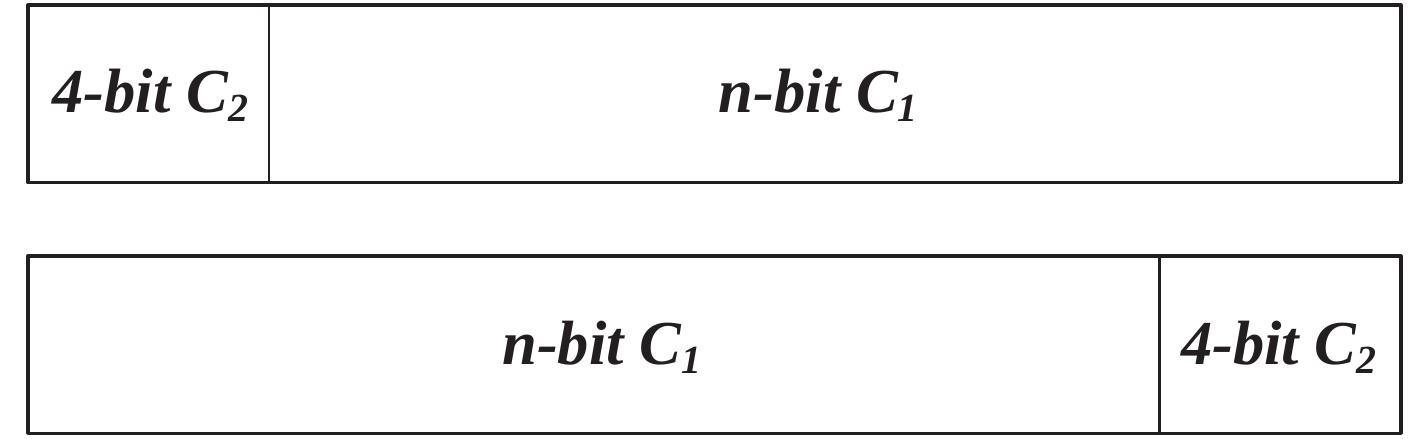}
\caption{The schematic diagram of external challenge block segmentation}\label{fig31}
\end{figure}

\begin{figure}[!t]
\centering
\includegraphics[width=2.5in]{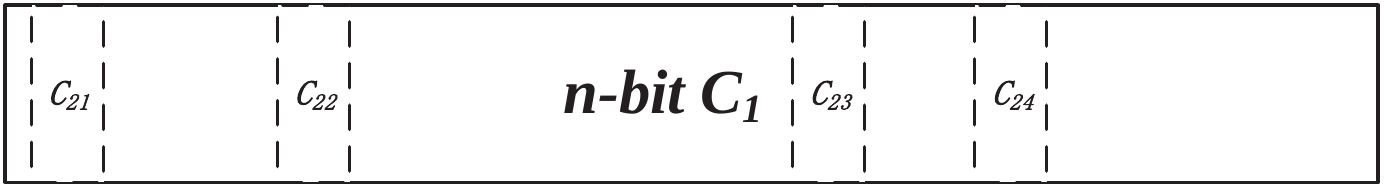}
\caption{The schematic diagram of external challenge scattered segmentation}\label{fig32}
\end{figure}

In the proposed design scheme, $(n+4)$-bit $C$ is categorized
into two parts. In the first part, $n$-bit $C_1$ is utilized as the {LFSR}'s
initial value, while in the second part, 4-bit $C_2$ determines the
shift count of the {LFSR} and applies it to the Control\;Unit. As
shown in Fig. \ref{fig31} and Fig. \ref{fig32}, these two parts can be roughly divided into the following segmentations:\par

\textbf{1) Block segmentation} refers to setting 4-bit $C_2$ as a block at
the beginning, end, or middle of $(n+4)$-bit $C$ (see Fig. 3
where the $C_2$ block is located at the starting position of the $C$).\par

\textbf{2) Scattered segmentation} refers to scattering 4-bit $C_2$
within $(n+4)$-bit $C$. As shown in Fig. \ref{fig32}, $C_{21}$, $C_{22}$, $C_{23}$, and $C_{24}$,
randomly embedded in n-bit $C_1$, form the 4-bit $C_2$, respectively.\par

Different external challenges can be employed to design an
obfuscation scheme according to different division strategies
for different systems and applications. In this paper, the more concise segmentation method, i.e.\,, the block segmentation method is employed.\par

The value of $4$-bit $C_2$ determines the shift count of the $n$-stage
$LFSR$.\,  For simplicity, the value of $C_2$ is chosen to be equal to the $Shift\;
Count$ of the $LFSR$. Thus, there are 16
shift count possibilities. A base integer corresponding to the $Shift\;Count$ is
predefined to further enhance the modeling attack resistance of
the obfuscation scheme with the $LFSR$. Now, the final $Shift\;
Count$ can be formulated as follows:

\begin{equation}
Shift\;Count = C_2*Base \label{eq77}
\end{equation}

where $C_2$ is a 4-bit binary number, $Base$ is a predefined
integer value, and $Shift\;Count$ is the ultimate number of $LFSR$
shifting controlled by shift signal $S_L$. If the $Base$ is equal to 100,
the $Shift\;Count$ should be increased to 100 times the value of
$C_2$ after each external challenge input.\par

$LFSR$ includes a shift register and a linear feedback function
employed as a pseudo-random sequence generator. The primary input of the $LFSR$ is called the seed, while the bits
affecting the $LFSR$’s next state are called taps. $LFSR$ can be
divided into $Fibonacci \; LFSR$ and $Galois \; LFSR$ according to
different feedback methods \cite{bib15}. Since the hardware
implementation of the $Galois \; LFSR$ is generally faster than that of the
$Fibonacci \; LFSR$ due to the reduction in the number of logic
gates in its feedback loop, the former is chosen in this work.\par

A time-variant obfuscation scheme is proposed here. Several bits are extracted from the external challenge and applied to the mapping before the external challenge obfuscation. The specific external challenge in real-time is then used to determine the mapping of the obfuscation module. This is equivalent to changing the fixed mapping into a mapping varying with the external challenge and transforming the obfuscation mechanism from “time-invariant” to “time-variant,” which significantly reduces the attacker’s ability to break it by modeling attack. Moreover, the {LFSR}-based obfuscation logic can be easily updated by changing the tap sequence of the {LFSR} to resist the replay attack.

\begin{figure}[!t]
\centering
\includegraphics[width=3.5in]{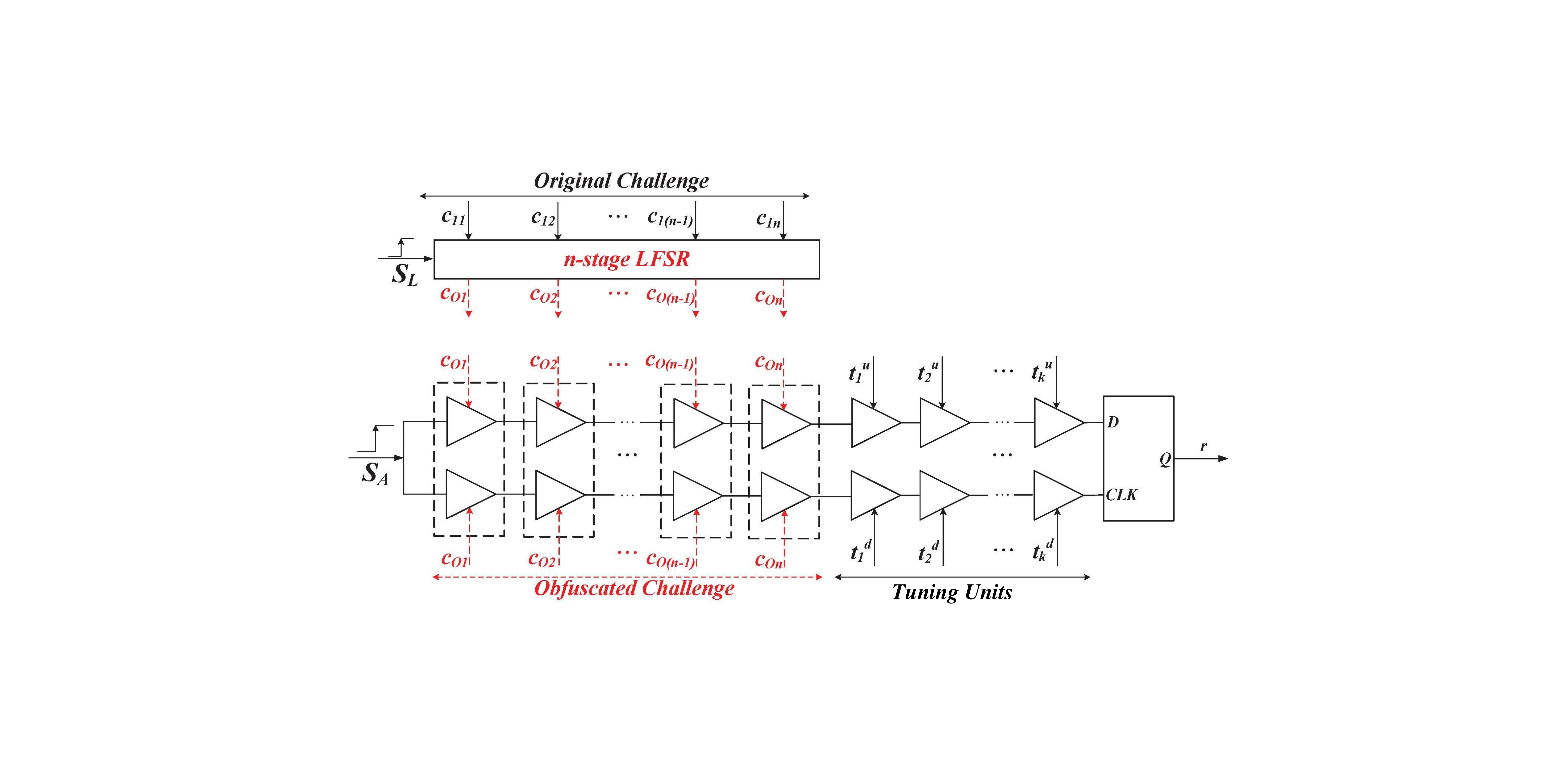}
\caption{The PDL-based LFSR-PUF structure}\label{fig4}
\end{figure}

\subsection{Implementation of the Proposed LFSR-APUF on FPGA}\label{subsec33}

Although various original and improved PUF structures have been proposed in the literature, some issues regarding PUF implementation, especially implementation on FPGA, warrant further investigation. The APUF operates by comparing the difference between two delay chains with a similar nominal delay based on the design process but different accurate delays generated by the construction process. However, the placement and routing of APUF are confined by the internal structure of FPGA as a kind of semi-custom integrated circuit,\,  leading to the severe asymmetry of APUF implemented on FPGA.\par

The proposed {LFSR-APUF} is implemented on 
 Xilinx Artix-7 FPGA. FPGA provides a generic substrate of interconnected blocks that can be (re)programmed several times. Two independent programmable delay lines (PDLs) with low overhead are utilized to introduce a non-path-swapping switch
block for implementing APUF, alleviating the system delay
bias of the two delay chains as much as possible \cite{bib16},\cite{bib17}. A single the lookup table (LUT) of FPGA is employed for PDL
implementation.\, 
By changing the signal input to the LUT, the signal propagation path of the LUT can be adjusted to affect the signal propagation delay of the PDL. 
 As shown in Fig. \ref{fig4}, 
the \textbf{\emph{Original Challenge}} is the $n$-$bit$ $C_1 (i.e.\,c_{11}, c_{12} \ldots , c_{1(n-1)}, c_{1n})$  mentioned in Fig. \ref{fig2},
and the \textbf{\emph{Obfuscated Challenge}} is the $n$-$bit$ $C_O (i.e.\,c_{O1}, c_{O2} \ldots , c_{O(n-1)}, c_{On})$  mentioned in Fig. \ref{fig2}.
The LFSR-APUF circuit contains n new non-path-swapping switch units and k tuning
units. The upper and lower PDL pairs form a PDL block,
which can be divided into two types: switching block and tuning block. The
two PDLs of the switching block are applied to the same $c_{Oi}$ (i = 1, 2, \ldots , {n-1}, n), 
while the two PDLs of the tuning block are applied to
the tuning bits $t_i^u$ and $t_i^d$ (i = 1, 2, \ldots, k) as independent signals
controlled by the user. The challenge controls the switching
block, while the user directly controls the tuning block by
incorporating an appropriate additional delay to the top or
bottom delay chain to counteract the delay deviations caused by
asymmetric routing. 
If the measurement result shows that there are more '1' in the APUF instance responses, it indicates that the top delay chain is too short, then we need to appropriately extend the top delay path by changing the corresponding tuning bit; otherwise, the bottom delay path needs to be appropriately extended if there are more '0'.
\, The users generally adjust the tuning bits
so that the measured uniformity approaches 50\%, while the
tuning bits can be fixed. After the tuning bit is fixed, attackers cannot change it.\par

Since APUF and obfuscation logic are relatively
independent in the proposed {LFSR-APUF} design, it is necessary
to \emph{Synthesize} and \emph{Place} \& \emph{Route} again after modifying the
obfuscation logic by behavioral-level hardware
description language (HDL). This leads to unacceptable results
influencing the original routing of the APUF circuit. The
best way to solve the above problem is to implement an APUF
hard macro. The following three steps should be performed to
design an unbiased APUF hard macro.\par

1) Designing a single PDL hard macro.\par

2) Designing a PDL chain hard macro.\par

3) Designing an arbiter PUF hard macro.\par

\begin{figure}[!t]
\centering
\includegraphics[width=\columnwidth]{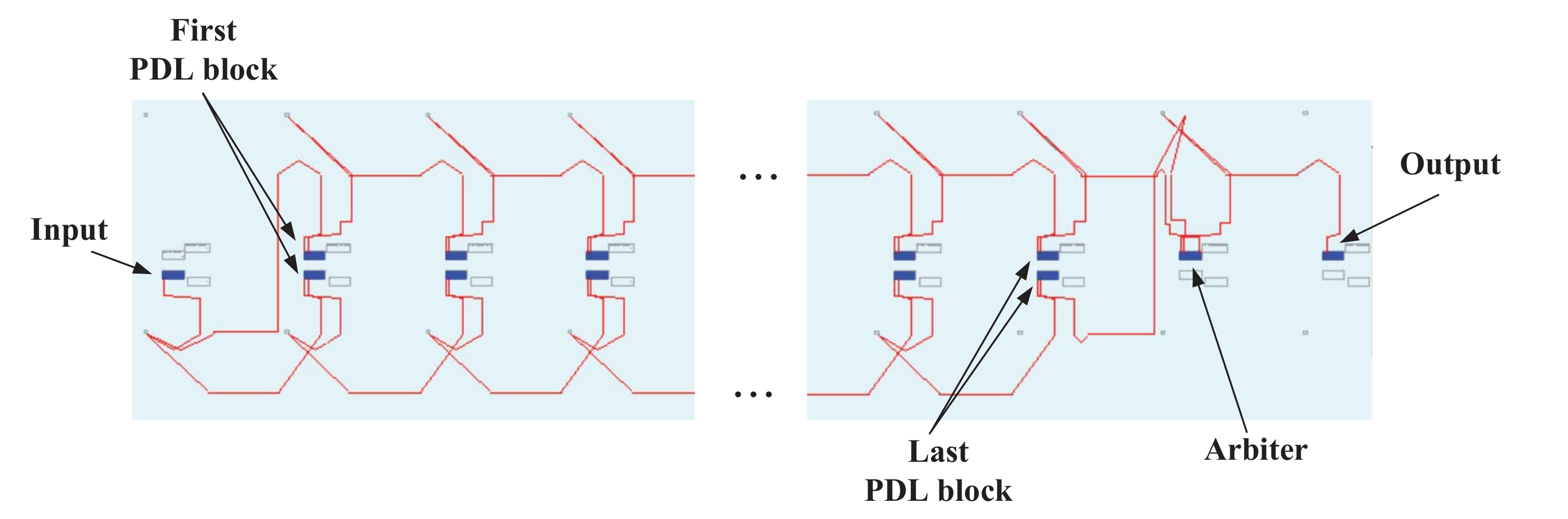}
\caption{Placement and routing of the APUF hard macro}\label{fig5}
\end{figure}

The specific placement and routing of the APUF hard
macro are presented in Fig. \ref{fig5}.\,  The routing before the first PDL
block and between the last switching block and arbiter (D flip-flop)
is asymmetric, while the rest is entirely symmetric. The
APUF hard macro can be directly represented as the
bottom module to implement the proposed time-variant
obfuscation scheme or other schemes without modification.

\section{Experiment Results of the Proposed LFSR-APUF}\label{sec4}

The proposed LFSR-APUF is implemented on five {Xilinx Artix-7 XC7A100T} FPGAs. The received experimental data is
processed using Python language. Specifically, the onboard  {UART} interface of PC is used to transmit the challenge to the FPGA and return the response. The PC replaces the
\emph{Challenge Generator} mentioned in Section \ref{sec3} to generate
random external challenges.\par

In this section, the concept of \textbf{\emph{Hamming distance}} \textbf{(HD)} is used repeatedly. For the m-bit bitstream X and Y, the \textbf{HD} of X and Y is defined as follows,

\begin{equation}
   \emph{HD}(X,Y) = 
\frac{1}{m}\Biggl(\sum_{i=0}^{m-1}X[i] \oplus Y[i]\Biggl) \times 100\% \label{eq8}
\end{equation}

\subsection{Resistance to Modeling Attack}\label{subsec41}

\begin{figure}[!t]
\centering
\includegraphics[width=2.5in]{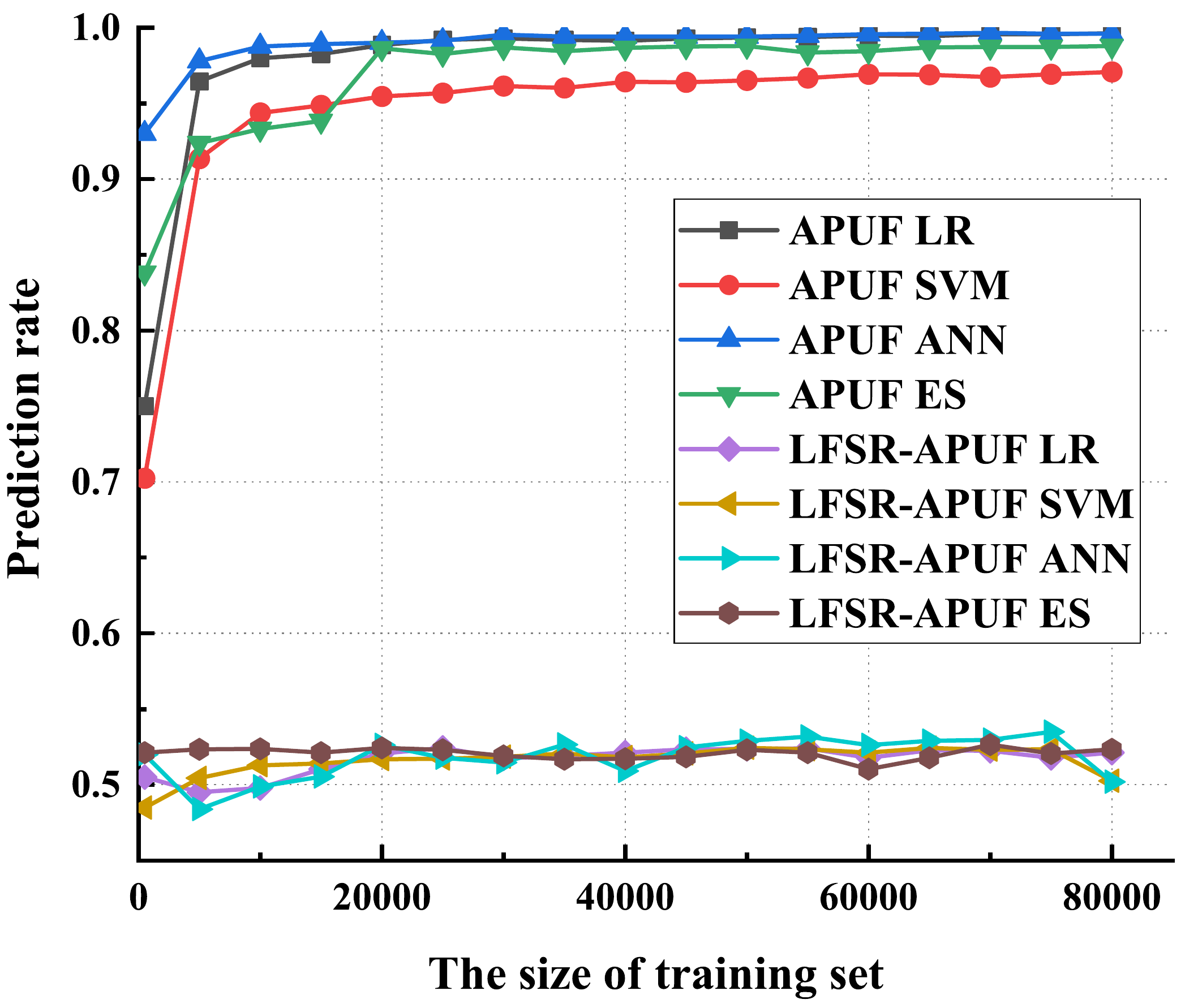}
\caption{Result of modeling attacks on arbiter PUF and LFSR-APUF}\label{fig6}
\end{figure}

Eighty thousand CRPs are collected for APUF, and the
proposed {LFSR-APUF} is implemented on a {Xilinx Artix-7} FPGA board to
verify the modeling attack resistance of the APUF. 
Both circuits consist of 64-stage PDL blocks.
Mainly, LR, ES, ANN, and SVM are utilized to evaluate the
circuits. Here, 80\% of the dataset is devoted to training,
while the remaining 20\% is devoted to testing. Fig. \ref{fig6} presents
the prediction rate of modeling attacks for 64-stage APUF
and 64-stage $LFSR-APUF$ under different training CRP values.\par

\begin{figure}[!t]
\centering
\includegraphics[width=2.5in]{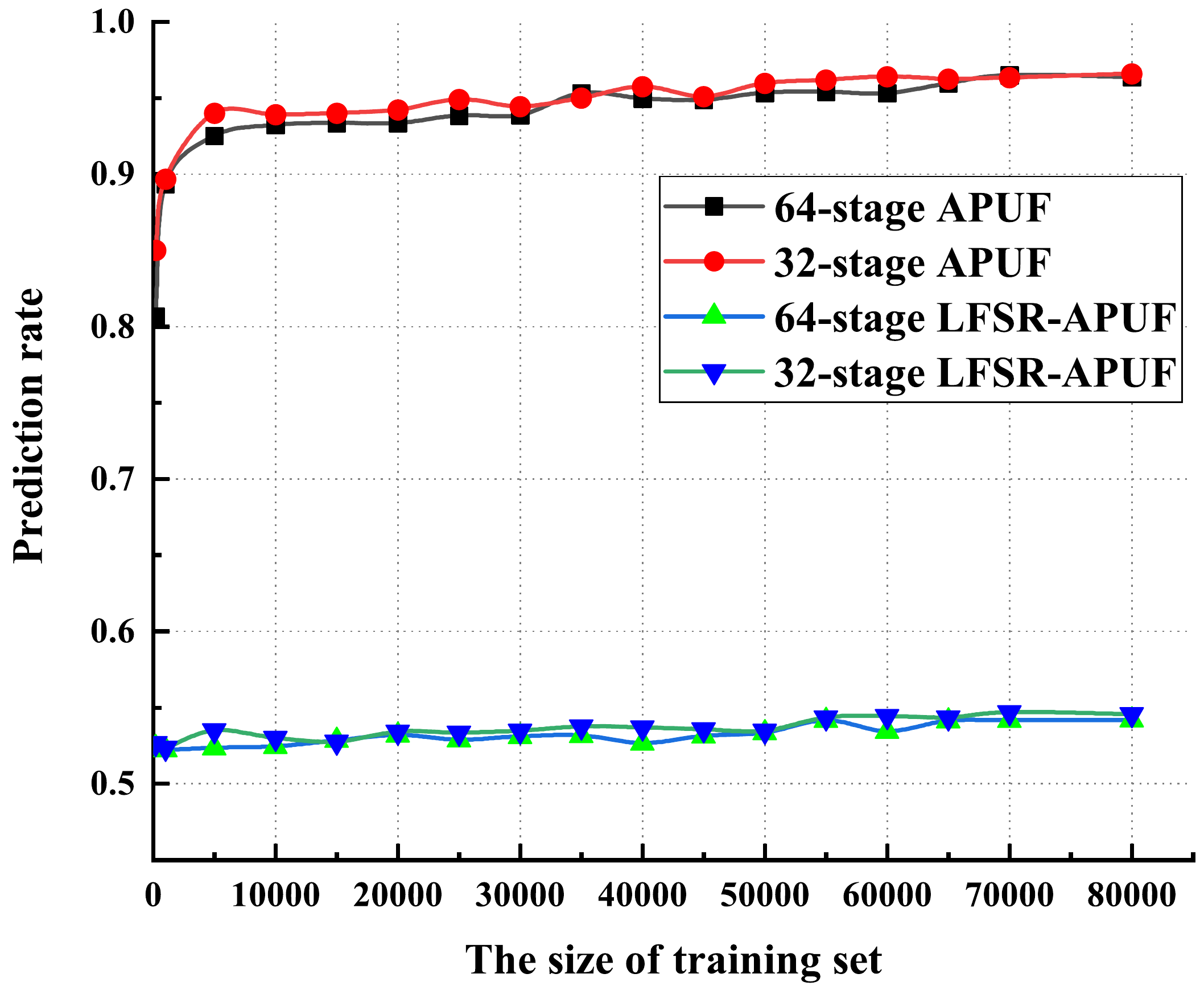}
\caption{Result of LR attack on arbiter PUF and LFSR-APUF with 32 and 64 stages}\label{fig7}
\end{figure}

For simplicity, the $Base$ value is chosen as 1. As shown in Fig. \ref{fig6}, the APUF can still be predicted effectively, even if there are fewer than 100 training data points, while the modeling attack prediction rate is as high as 85\%. When the training set size is more than 2000, the prediction rate of the modeling attack is 91.4\% $\sim$ 99.1\%. However, since the attacker can only obtain the original external challenge and not the internal obfuscated challenge, the prediction rates of the four attack algorithms for the proposed {LFSR-APUF} are maintained at 48\% $\sim$ 54\%.\ Furthermore, the LR algorithm is utilized for modeling the 32-stage APUF and 32-stage {LFSR-APUF} to assess the resistance of {LFSR-APUF} to the modeling attack under stage variations. Fig. \ref{fig7} compares their prediction rates with those of 64-stage APUF and 64-stage {LFSR-APUF}. The prediction rate of {LFSR-APUF} is much lower than that of APUF for both 32 and 64 stages. Notably, the 32-stage {LFSR-APUF} has a slightly higher prediction rate than the 64-stage one. Consequently, the obfuscation logic significantly improves the modeling resistance for both 32-stage and 64-stage APUFs.\par

\begin{figure}[!t]
\centering
\includegraphics[width=2.5in]{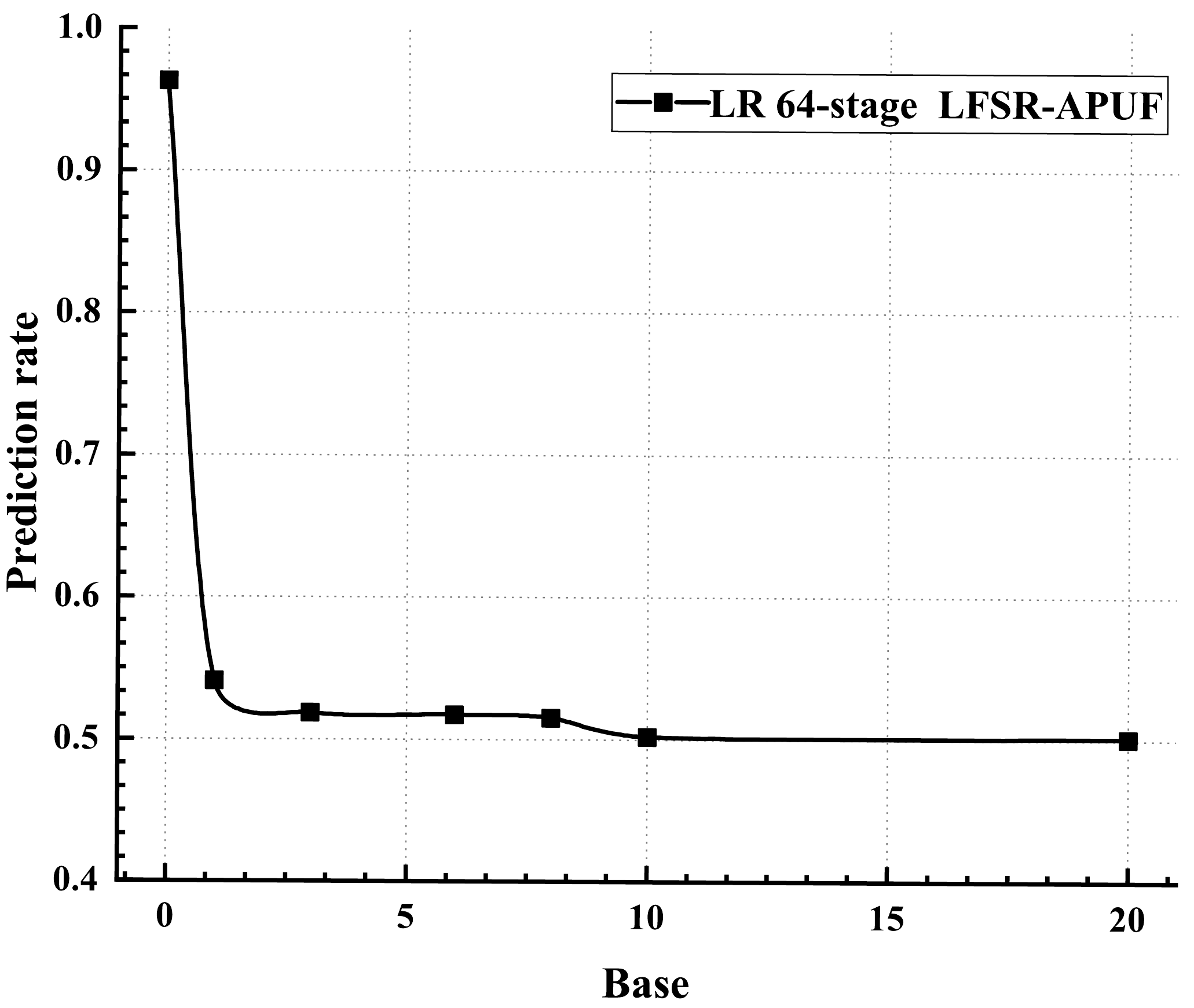}
\caption{Prediction rate of 64-stage LFSR-APUF with various \emph{Base} values}\label{fig8}
\end{figure}

Different \emph{Base} values (0, 1, 3, 5, 8, 10, and 20) are chosen in Equation \ref{eq77} to further discuss its influence on the resistance to the modeling attack. Besides, 10,000 CRPs are collected from the proposed 64-stage {LFSR-APUF} for modeling with LR, where the prediction rates are presented in Fig. \ref{fig8}. When the \emph{Base} is 0, the LFSR does not affect the prediction rate, and the prediction rate is approximately 93\%; when the \emph{Base} is 1, the prediction rate significantly decreases to 54\%, demonstrating that the obfuscation mechanism can increase the resistance to modeling attack. When \emph{Base} increases to 10, the prediction rate gradually decreases to nearly 51.79\%, almost equivalent to random guess. Therefore, it is recommended to choose the \emph{Base} value as 10 in practical applications.

\begin{table}[!t]
\caption{$LFSR-APUF$ $Evaluation$ $Metrcs$}\label{tab1}
\setlength{\tabcolsep}{9mm}
\centering
\begin{tabular}{@{}ccccc@{}}
\hline
$Property$\; & $ Ideal $  & $LFSR-APUF$  \\   
\hline
$Ramdomness$ & 50 \%{}  & 51.35 \%{}  \\
$Reliability$ & 100 \%{} &  99.3 \%{} \\
$Uniqueness$ & 50 \%{}  & 49.1 \%{} \\
\hline
\end{tabular}
\end{table}

\subsection{Randomness}\label{subsec42}

 Randomness is generally employed to measure the distribution
of '0' and '1' in the PUF responses under different input
challenges.  Randomness is defined as follows,

\begin{equation}
 Randomness = (\frac{1}{m}\sum_{i=1}^m R_i)\times 100\% \label{eq9}
\end{equation}

Where, $m$ represents the number of PUF response bits, 
and $R_i$ represents the $i-th$ bit response of the PUF instance. Table I shows  various evaluation metrics of  LFSR-APUF.
If the response bit stream of a PUF is statistically random enough, 
its randomness should be ideal, i.e.\,, 50\%.
In our experiment, 80,000 challenges are randomly
generated, and the input to the {LFSR-APUF} is repeated ten times. 
Finally, the average  randomness is approximately obtained as
51.35\%, which is close to the desired value.

\subsection{Reliability}\label{subsuc43}

Reliability is generally adopted to evaluate the
reproducibility  of response in different environments of the same
challenges, which can be obtained by measuring the intra-chip
\textbf{HD} of PUF. The Reliability is defined as follows:

\begin{equation}
Reliability = (1-\frac{1}{m}\sum_{i=1}^m \frac{HD(R,R_i)}{n})\times 100\% \label{eq10}
\end{equation}

where $m$ represents the number of tests, $n$ represents the sequence length of PUF response, $R$ represents the reference response value, and $R_i$ represents the value of $i-th$ response under different tests. Ideally, if the same challenges are applied to the same PUF in different environments, the PUF should generate the same response, i.e.\,, the ideal value of reliability is 100\%. Reliability is a severe problem for FPGA-based PUF \cite{bib18},\cite{bib19},\cite{bib20}. A voting module and a soft dark-bit masking module\cite{bib34} are introduced in the underlying APUF to further improve the reliability.\ In our measurement, 80,000 challenges are randomly generated and applied to {LFSR-APUF} 20 times respectively at -20 ℃, 0 ℃, 20 ℃, 25 ℃, 30 ℃, 50 ℃ and 70 ℃, respectively. As shown in Table I, the average reliability is calculated as 99.3\%. 

\subsection{Uniqueness}\label{subsuc44}

For the same PUF design, if the same challenge is given, the difference among the responses generated by different PUF instances should be as large as possible, i.e.\,, there should be a certain degree of differentiation. From a mathematical point of view, the different instances of the same PUF design should be statistically independent and random. The uniqueness of PUF can be obtained by measuring its inter-chip \textbf{HD}. Uniqueness is defined as follows:

\begin{equation}
Uniqueness = (1-\frac{2}{k(k-1)}\sum_{i=1}^{k-1} \sum_{j=i+1}^{k} \frac{HD(R_i,R_j)}{n})\times 100\% \label{eq11}
\end{equation}

Where $k$ represents the number of PUF instances tested, 
$R_i$ and $R_j$ represent the response generated by the $i-th$ instance and $j-th$ instance, respectively, under the same test conditions, 
and $n$ represents the sequence length of all the responses of the PUF instances. Ideally, the uniqueness of PUF should be 50\%.
In our measurement, 80,000 challenges are randomly generated, and
$LFSR-APUF$ is then implemented on five FPGAs. Accordingly,
the measured mean uniqueness is obtained as 49.1\%.

\subsection{Hardware Overhead}\label{subsuc45}

\begin{table}[!t]
\caption{$Comparison$ $about$ $hardware\;overhead$ $with$ $other\;PUF$}\label{tab2}
\centering
\begin{tabular}{@{}ccccc@{}}
\hline
$Design$ & $Challenge(bit)$ & $Slice\;LUT$  & $Slice\;Register$  \\   
\hline
$LFSR-APUF$ & 64 & 383 & 399 \\
$PUF-FSM$ \cite{bib32}& 128 & 960 & 1500 \\
$RPUF$\cite{bib32}  & 128 & 350 & 389 \\
$RSO-PUF$ \cite{bib8} & 128 & 395 & 176 \\
\hline
\end{tabular}
\end{table}

As shown in Fig. \ref{fig4}, the hardware overhead of {LFSR-APUF} is low as it only contains one {LFSR} and some necessary control logic circuits more than that in traditional APUF. The {LFSR} consists of n D-flip flops and several XOR gates, where n indicates the number of challenge bits and the number of XOR gates, which depends on the number of taps. For the {LFSR-APUF} with the 68-bit challenge  (including 64-bit $C_1$ and 4-bit $C_2$, as shown in Fig. 2)  and 1-bit response, the numbers of Slice LUTs and registers occupied by the 64-stage LFSR-APUF are 383 and 399, respectively. Compared to the baseline APUF, the extra overhead of {LFSR-APUF} is 18 \% of the total overhead. A comparison between {LFSR-APUF} with other MLA-resistant PUFs is presented in Table \ref{tab2}. Although the hardware overhead of {LFSR-APUF} is higher than that of {RPUF} and {RSO-PUF}, the  MLA-resistant ability of {RPUF} is lower than that of {LFSR-APUF}, and {RSO-PUF} requires an additional consumption of 4KB of RAM.

\section{Proposed PUF-Based Authentication Protocol}\label{sec5}

The proposed protocol consists of a registration phase and an authentication phase. During the registration phase, each IoT device $Device_k$ needs to be registered on the server. In the authentication phase, mutual authentication is required between a device and a server, i.e.\,, each device should prove its authenticity to the server and the server should also prove its legitimacy to the device.

\subsection{Adversary Model}\label{subsuc51}

Since the security of the same authentication protocol is different under different adversary models, we first need to define the adversary capabilities through the adversary model to validate the security of the proposed authentication protocol. \par

1) The attacker is allowed to eavesdrop, replay, manipulate, and block information traffic on all communication links between the server and the device as well as between the server and the database during the authentication phase.\par
 
2) Based on all information traffic among the device, server, and database obtained, the attacker can summarize, deduce and analyze them to find some repetition rules and then perform man-in-the-middle attack and spoofing attack to obtain illegal access.\par

3) Cloning attacks and intrusive physical attacks are allowed on the devices. Further, the attacker is allowed to conduct MLAs on the device, enabling him/her to make brute force queries on the device through an open interface to predict unknown session information. \par

4) The server and database are in an insecure environment. An attacker can take full control and attack the server to gain internal information, and he/she also has free access to the information stored in the database.\par

\begin{figure*}[!t]
\centering
\includegraphics[width=5.5in]{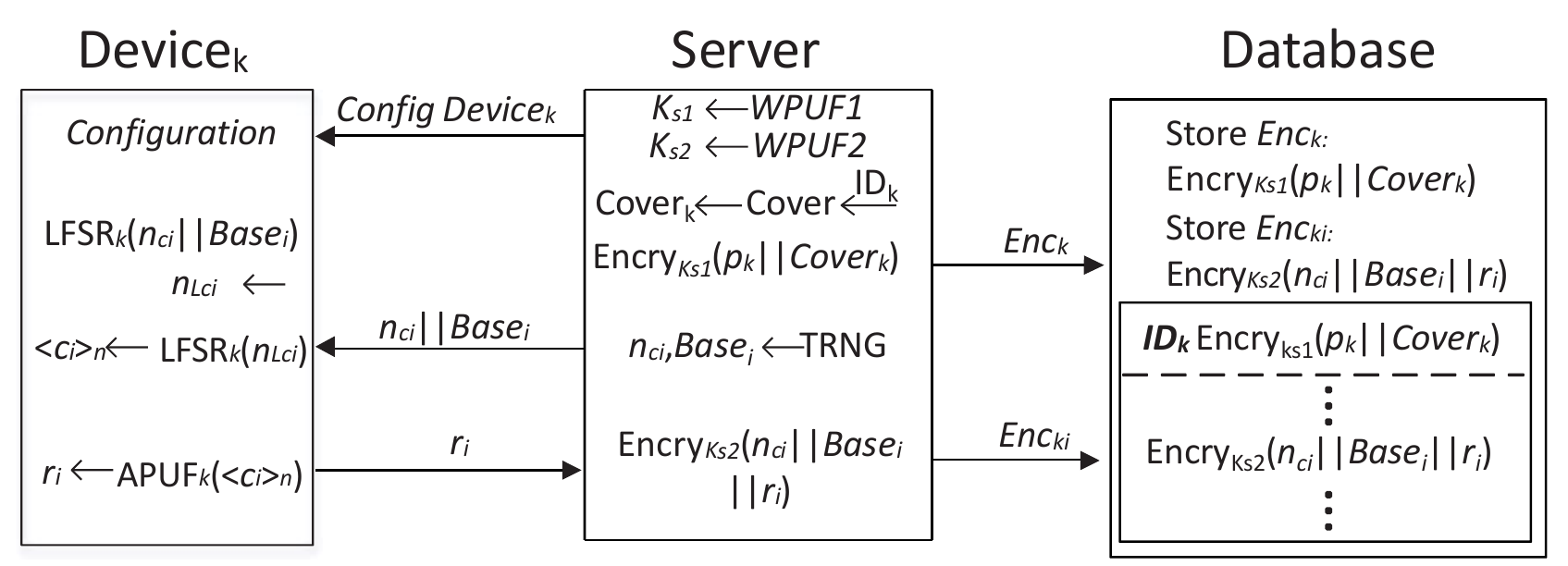}
\caption{Registration Phase}\label{fig9}
\end{figure*}

Before the registration phase, the following assumptions are made: the network environment of the proposed authentication protocol includes mutual authentication between the server and multiple devices. Upon leaving the factory, the server should contain two WPUFs, indicated as $WPUF_1$ and $WPUF_2$, which can produce a unique and stable response once powered on. In addition, there are $x$ devices in the authentication network. A unique APUF should be embedded inside each device. For device $k$, it is defined as $APUF_k$, and each APUF is unique, random, and stable.\par

\subsection{Registration Phase}\label{subsuc52}

The detailed steps of the registration phase are shown in Fig. \ref{fig9}. These steps are described as follows:\par

Step1 : \emph{WPUF}$_1$ and \emph{WPUF}$_2$ generate two server keys after power-on, 
which are defined as $K_{s1}$ and $K_{s2}$, respectively.\par

Step2 : The server assigns a unique identity to each device in the authentication network. 
The unique identity of $Device_k$ is defined as $ID_k$. 
Similarly, the server assigns each device a unique primitive polynomial for the device-specific {LFSR} design, 
and the primitive polynomial for $Device_k$ is defined as $p_k$. 
The server configures $Cover$ with $ID_k$ to generate the $Cover_k$, which is a unique ultra-lightweight $Cover$ function for each device.
\par

Step3 : The server encrypts the concatenation of $p_k$ and $Cover_k$ with the encryption key $K_{s1}$, 
defined as $Encry_{K_{s1}} (p_k \| Cover_k)$ and abbreviated as $Enc_k$, 
in which the symbol $\|$  is used to represent the concatenation of two vectors. 
Then $Encry_{K_{s1}} (p_k \| Cover_k)$ is stored in the database with $ID_k$ as the index, 
and some space should be reserved under this index 
for storing data related to $APUF_k$.
Note that each device in the authentication network is assigned a partition 
in the database to store relevant information,\, as shown in Database in Fig. \ref{fig9}.\par 

Step4 : A true random number generator $TRNG$ in the server 
generates $y$ sets of random numbers $(n_c,\;Base)$. 
For the round $i$ of authentication  $Device_k$, 
$(n_c,\;Base)$ are defined as $(n_{ci},\;Base_i)$. 
$n_{ci}$ is $n$-bit long, equivalent to APUF circuit stage, 
and $y$ determines the size of the CRP dataset used for authentication with the corresponding device. The server removes the $n_{ci}$ whose $C_1$ is all-zeros or all-ones to prevent the {LFSR} from getting stuck.\par

Step5 : For the $i$-$th$ authentication round of $Device_k$, 
the received $(n_{ci} \| Base_i)$ is input into $LFSR_k$, 
which determines the seed value and shift count value of $i$-$th$ authentication round for $LFSR_k$, 
to generate the corresponding obfuscated challenge $n_{Lci}$.\par

Step6 : $LFSR_k$ in $Device_k$ takes the $n$-bit $n_{Lci}$ 
generated in the previous step as the seed, named as the initial challenge, to generate $n$-bit sub-challenge, 
which is defined as $<c_i>_n$, and then $<c_i>_n$ is sent to the underlying $APUF_k$.\par

Step7 : $APUF_k$ in $Device_k$ uses the received $<c_i>_n$ to generate an $n$-bit response, 
defined as $r_i$, which is sent to the server. The server checks whether the CRP is reliable.\par

Step8 : The server encrypts the reliable ($n_{ci} \| Base_i \| r_i$) with the encryption key $K_{s2}$, 
which is defined as $Encry_{K_{s2}}$$ (n_{ci} \| Base_i \| r_i)$, abbreviated as $Enc_{ki}$, 
and stores it in the reserved space under the $ID_k$ index in the database.

\subsection{Authentication Phase}\label{subsuc63}

\begin{figure*}[!t]
\centering
\includegraphics[width=6.0in]{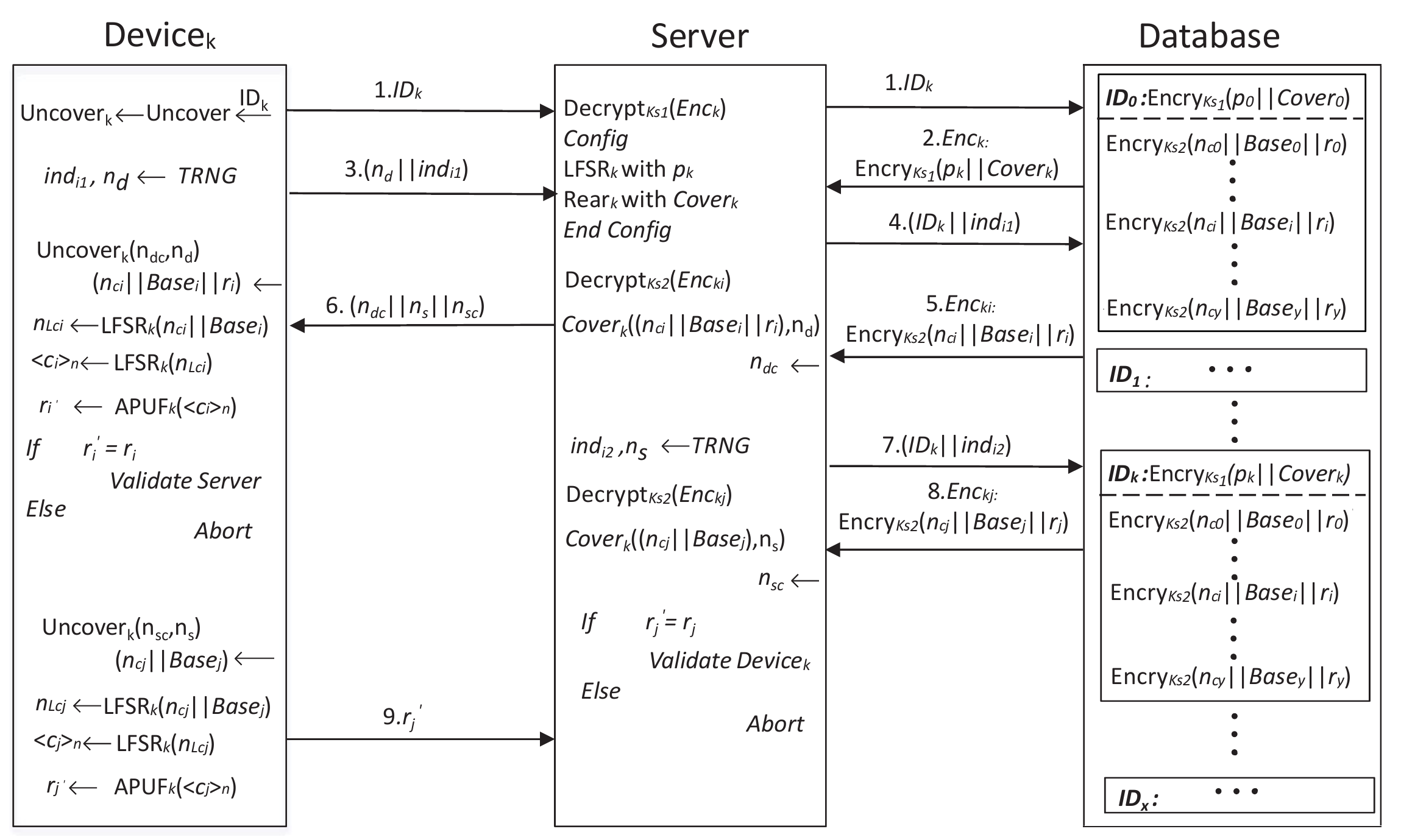}
\caption{Authentication Phase}\label{fig10}
\end{figure*}

After all the legitimate devices are registered on an authentication network, the network enters the authentication phase. Unlike the registration phase which occurs in a secure environment, the entire network is in an insecure environment during the authentication phase. The authentication phase with the $i-th$ authentication round of $Device_k$ is shown in Fig.\ref{fig10}. This phase includes the following steps: \par

Step1 : $Device_k$ sends its own $ID_k$ for the server to initial authentication communication.
After receiving the $ID_k$ from $Device_k$, 
the server simply checks its validity and sends it to the database 
for requesting the encrypted message related to $Device_k$. By the way, the $Device_k$ configures $Cover$ with $ID_k$ and makes it become a unique $Cover_k$.
\par

Step2 : After receiving the $Enc_k$ about $Device_k$ from the database, 
the server uses the key $K_{s1}$ generated by \emph{WPUF}$_1$ to decrypt it,  
generating $(p_k \| Cover_k)$, and then the server not only uses the primitive polynomial $p_k$ to configure $LFSR_k$, 
but also uses $Cover_k$ to configure the private bit conversion function.
Note that both $LFSR_k$ and $Cover_k$ are unique and are bound to the $Device_k$.\par

Step3 : When the server completes the above configuration, 
$Device_k$ uses its own TRNG to generate $(n_d, ind_{i1})$, 
where $n_d$ is a random $l$-bit for $Cover_k$, 
and $ind_{i1}$ is used to request $i$-$th$ round encryption message in the database.\par

Step4 : After receiving $(n_d, ind_{i1})$ from the device, 
the server uses $(ID_k, ind_{i1})$ to request $Enc_{ki}$ from the database. 
The database then sends the $i-th$-round encryption message $Enc_{ki}$ about $Device_k$ to the server.\par

Step5 : After receiving the encrypted message from the database, 
the server decrypts it with the key $K_{s2}$ generated by \emph{WPUF}$_2$, 
to produce $(n_{ci} \| Base_i \| r_i)$.
The server then  \textbf{\emph{Covers}} $(n_{ci} \| Base_i \| r_i,\;n_d)$ , resulting in  $n_{dc}$.
Finally, the server sends the $n_{dc}$ to $Device_k$.\par

Step6 : The $Device_k$ \textbf{\emph{Uncovers}} the received $n_{dc}$ to obtain the relevant entry, namely $(n_{ci} \| Base_i \| r_i)$.
The $Device_k$ then sends the received $(n_{ci} \| Base_i)$ to $LFSR_k$ for the obfuscated challenge $n_{Lci}$, 
and subsequently $n_{Lci}$ is used as the seed of $LFSR_k$ to obtain $n$ $n$-bit sub-challenge $<c_i>_n$. 
Finally, $<c_i>_n$ is sent to \emph{APUF}$_k$ to generate an $n$-bit response $r_i$.
Because the server extracts information about the session during the registration phase, 
the $r_i$ sent by the legitimate server should be the same as the $r_i'$ generated by $Device_k$.
If the $r_i$ and $r_i'$ are the same, the device validates the server.\par

Step7 : After the server is validated, the server uses its own TRNG to generate $(n_s, ind_{i2})$ similar to Step3.
Then, the server sends the $n_s$ to $Device_k$, and on the other hand, 
it also sends $ind_{i2}$ to the database to request the related encryption message.\par

Step8 : After the server receives the $Enc_{kj}$ from the database, 
it decrypts it with the key $K_{s2}$, generating $(n_{cj} \| Base_j \| r_j)$.
The server then \textbf{\emph{Covers}} the $(n_{cj} \| Base_j)$ with $n_s$ to generate $n_{sc}$,
and sends it to $Device_k$.\par

Step9 :  Once $Device_k$ receives $n_s$ and $n_{sc}$ successively from the server, 
it \textbf{\emph{Uncovers}} the $n_{sc}$ with $n_s$ to obtain $(n_{cj} \| Base_j)$.
Next, $Device_k$ enters $(n_{cj} \| Base_j)$ into {LFSR-APUF} to generate $r_j'$, 
which is then sent to the server.
After receiving the $r_j'$ sent by $Device_k$, 
the server compares it with the $r_j$ extracted from the database.
If both are equal, the server validates the $Device_k$.
In this case, the server and the device have completed mutual authentication.

\subsection{Bit Conversion Through $Cover_k$}\label{subsuc54}

The bit conversion function $Cover_k$ has two $l-bit$ arguments. 
We define $l$ as 128, and define $t$ as 10.
The operation process of bit conversion function $Cover_k$ is mainly divided into three stages: 
bit rearrangement conversion process, parity adjacent cross XOR conversion process, 
and bit random filling conversion process.
In particular, different registered devices have different $Cover_k$ functions, 
 i.e.\,, the bit conversion function $Cover_k$ is unique for $Device_k$.\par

Next, we define the detailed usage of $Cover_k$ with $X,Y,Z,W,F$ and $O$. 
$X,Y,Z$ and $W$, are $l-bit$ binary vectors, $F$ is a $t-bit$ binary vector, 
and $O$ is a $(l+t)-bit$ binary vector, defined as follows:

\begin{equation}
X=x_1, x_2, x_3, \dots, x_l; x_i\in \{ 0,1 \},i=1,2,\dots,l
\end{equation}

\begin{equation}
Y=y_1, y_2, y_3, \dots, y_l; y_i\in \{ 0,1 \},i=1,2,\dots,l
\end{equation}

\begin{equation}
Z=z_1, z_2, z_3, \dots, z_l; z_i\in \{ 0,1 \},i=1,2,\dots,l
\end{equation}

\begin{equation}
W=w_1, w_2, w_3, \dots, w_l; w_i\in \{ 0,1 \},i=1,2,\dots,l
\end{equation}

\begin{equation}
F=f_1, f_2, f_3, \dots, f_t; f_i\in \{ 0,1 \},i=1,2,\dots,t
\end{equation}

\begin{equation}
O=o_1,o_2,o_3,\dots, o_{(l+t)}; o_i\in \{ 0,1 \},i=1,2,\dots,(l+t)
\end{equation}

\subsubsection{Bit Rearrangement}\label{subsucsuc541}

The bit rearrangement conversion process uses $Per(X,Y)$ function \cite{bib27}, 
and the conversion steps are as follows:

Step1: Initialize $p_x$ to vector $X$ and $p_y$ to vector $Y$.\par
Step2: Pointers  $p_x$ and $p_y$ move bit-by-bit simultaneously 
from the most significant bit (MSB) to the least significant bit (LSB).
During pointer movement, if the bit $p_y$ points to 1, 
then the bit pointed by $p_x$ is copied into vector $Z$.\par
Step3: Since the vectors $X$ and $Y$ are both $l$ bits, the pointers $p_x$ and $p_y$ reach the LSB at the same time.
Similarly, pointers $p_x$ and $p_y$ are returned bit-by-bit from LSB to MSB at the same time, 
and if $p_y$ refers to bit 0, the bit referred by $p_x$ is copied into vector $Z$.
Note that the bits from $X$ are copied to $Z$ in ascending order.\par
Step4: The bit rearrangement conversion process ends when the pointers $p_x$ and $p_y$ return to MSB again.\par

\begin{figure}[!t]
\centering
\includegraphics[width=\columnwidth]{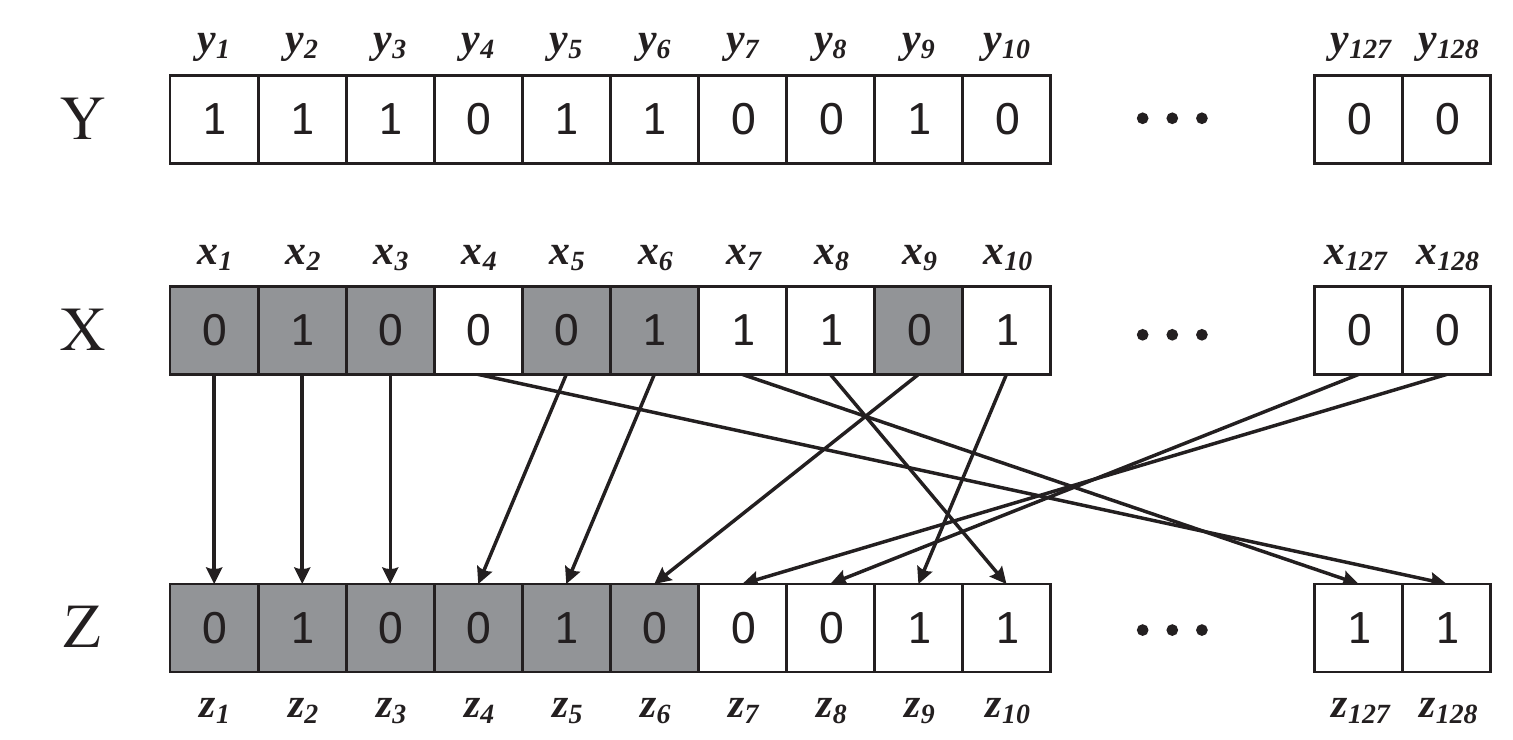}
\caption{Bit rearrangement conversion process}\label{fig11}
\end{figure}

The specific conversion process is shown in Fig. \ref{fig11}.
When $p_x$ and $p_y$ move from MSB to LSB at the same time, 
if the pointer $p_y$ refers to bit 1, 
the corresponding bit referred by the pointer $p_x$ in $X$ is represented by a \textbf{gray} background.
After $p_x$ and $p_y$  return from LSB to MSB at the same time, 
if the bit referred to by pointer $p_y$ is 0, 
the corresponding bit referred by pointer $p_x$ in $X$ is represented on a \textbf{white} background.

\subsubsection{Cross XOR}\label{subsucsuc542}
The parity adjacent cross XOR conversion process is as follows: \par

Step 1 : The result of bit rearrangement conversion process, namely vector 
$Z \; (Z \leftarrow Per(X,Y))$, needs to be extracted.\par
Step 2 : The adjacent odd and even bits in vector $Z$ and vector $Y$ are 
interchanged in pairs according to cross rule.\par
Step 3 : Every two adjacent bits in vector $Y$ are divided into a group, 
and then $l$-bit $Y$ can be divided into $l/2$ groups.
Next, two bits in every group need to be swapped. The swapped $Y$ is defined as $l$-bit vector $Y'$.\par
\begin{figure}[!t]
\centering
\includegraphics[width=\columnwidth]{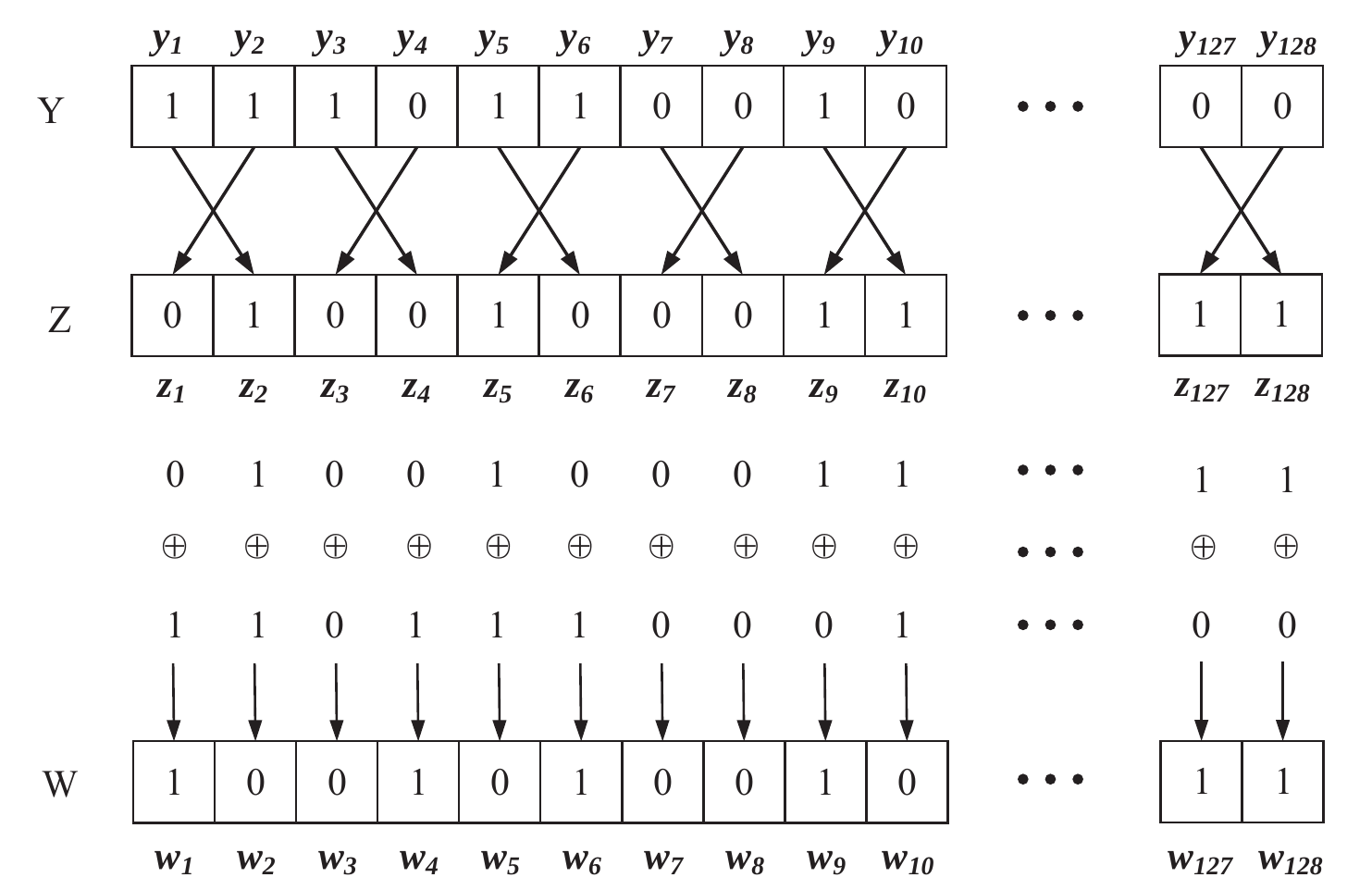}
\caption{Parity adjacent cross XOR conversion process}\label{fig12}
\end{figure}

Step 4 : The vector $Y'$ and vector $Z$ are bitwise XOR converted to obtain the $l$-bit vector $W$, as shown in Fig. \ref{fig12}.\par
In summary, in Equation \ref{eq18}, if $i$ is odd, then $w_i = y_i \oplus z_{i+1}$, otherwise $w_i = y_i \oplus z_{i-1}$,i.e.\,,

\begin{equation}
W=y_1 \oplus z_2 \| y_2 \oplus z_1 \| y_3 \oplus z_4 \| y_4 \oplus z_3 \dots y_{127} \oplus z_{128} \| y_{128} \oplus z_{127} \label{eq18}
\end{equation}

\subsubsection{Bit Filling}\label{subsucsuc543}

\begin{figure*}[!t]
\centering
\includegraphics[width=6in]{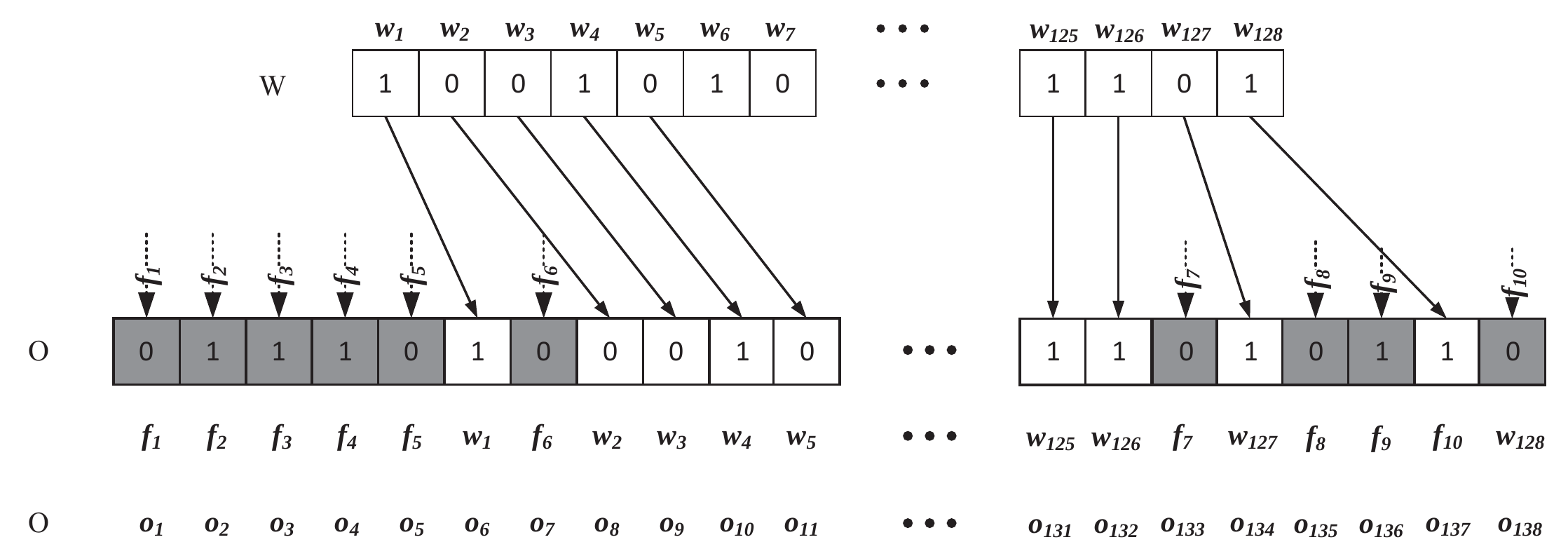}
\caption{Bit filling conversion process}\label{fig13}
\end{figure*}

The $Cover_k$ function defined in the previous section is used for bit filling. The $Cover_k$ function is uniquely bound to Devicek, i.e.\,, the $Cover$ functions are different for different devices. The uniqueness of the $Cover$ function is mainly reflected in the bit filling conversion process. First $t$-bit random vector $F$ is generated, which is then randomly filled into $l$-bit vector $W$. Obviously, there are $\binom{(l+t)}{t}$ bit-filling schemes. An example of the bit-filling scheme is shown in Fig.\ref{fig13}. According to the definition of $l$ and $t$ above, 
there are $\binom{(128+10)}{10} \approx 4.9 \times 10^{14}$ bit-filling schemes in total.
Such a large number of bit-filling schemes not only ensures that 
every registered device is assigned a unique $Cover$ function, 
but also makes it nearly impossible for an attacker to guess the $Cover_k$ of the embedded $Device_k$.
In addition, in every bit-filling scheme, there are $2^t$ possibilities in the confused filling bits 
to deceive the attacker, which further enhances the security of the $Cover_k$ function.

\section{Security Analysis and Comparison of the Proposed Authentication Protocol}\label{sec6}
In this section, we will analyze the different attack methods 
by which an attacker can compromise the proposed authentication protocol 
to demonstrate the security properties.

\subsection{Spoofing Attacks}\label{subsuc61}

It is assumed that an attacker has sufficient time and storage resources 
to monitor and store multiple authentication rounds between a server and several devices. 
Authentication rounds refer to the rounds in which all the legitimate devices and servers 
have completed mutual authentication after registration, 
including multiple authentication rounds between the same legitimate device and server, 
excluding the round in which legitimate devices have not completed mutual authentication with the server. 
After the above monitoring and storage process is completed, 
the attacker can impersonate the server, defined as $S_F$, 
and start a new round of authentication, namely $(i+1)-th$ round, with some legitimate device. 
In addition, the attacker can impersonate any device, defined as $D_{kF}$. 
For the sake of explanation, consider that the attacker has already monitored and stored 
the $(i+1)-th$ authentication round of $Device_k$.\par

The attacker impersonates the server to launch spoofing attacks as follows:\par

1) After receiving the $ID_k$, $n_d$ and $ind_{(i+1)1}$ from $Device_k$, 
$S_F$ starts to browse the stored $pre-i$ authentication round messages in its own database, 
to search for if there are exactly the same messages as $n_d$ and $ind_{(i +1)1}$
This is almost impossible. On the one hand, both $n_d$ and $ind_{(i+1)1}$ are generated 
by the true random number generator without any repetition rule. 
On the other hand, for $length(n_d)=138,\; length(ind_{(i+1)1})=138$, 
the probability of each random number occurring is negligible $1/2^{138} \approx 10^{-41}$.
Furthermore, the probability of two random numbers reappearing at the same time is even smaller.\par

2) Because $S_F$ cannot obtain the encryption keys $K_{s1}$ and $K_{s2}$, 
it cannot decrypt the encrypted messages from the database. 
Even if $Enc_k$ and $Enc_{k(i+1)}$ of the $(i+1)-th$ authentication round from $Device_k$ 
can be precisely obtained, $S_F$ cannot take any action.
$S_F$ can only select a message from the previous authentication rounds and send it to $Device_k$.
If $S_F$ sends $i-th$-round message $n_{dc}$ to $Device_k$, 
$Device_k$ uncovers the received $n_{dc}$ with $n_d$, generating sequence $(n_{cif} \| Base_{if} \| r_{if})$.
$Device_k$ then enters $(n_{cif} \| Base_{if})$ into $LFSR-APUF_k$ to generate $r_i'$.
Because $r_{if}$ cannot be the same as $r_i'$, legitimate $Device_k$ rejects the fake server $S_F$.\par

3) $S_F$ can also launch a spoofing attack on $Device_k$ with authentication round messages stored in the database 
from other devices, such as $Device_{k-1}$ and $Device_{k+1}$. This is obviously nearly impossible. 
If $S_F$ uses authentication round messages from other devices to spoof $Device_k$, 
the uniqueness and unclonability of the APUF embedded in the device must be considered.
Considering that the uniqueness of 64-stage {LFSR-APUF} measured in the experiment described in Section \ref{subsuc44} is 49.1\%, 
for {LFSR-APUF}$_{(k-1)}$, {LFSR-APUF}$_{k}$ and {LFSR-APUF}$_{(k+1)}$,
even if the same challenge is applied, 
the probability of the same responses generated by them is negligible.\par

The attacker impersonates $Device_k$ as $D_{k_F}$ to launch a spoofing attack:\par

4) $D_{k_F}$ initiates $(i+1)-th$ authentication round in the legitimate server.
After $D_{k_F}$ validates the legitimate server fraudulently, 
the server sends $n_s$ and $n_{sc}$ to $D_{k_F}$ to determine if $D_{k_F}$ is trustworthy. 
Similar to the above case in which legitimate $Device_k$ rejects the fake server $S_F$, 
the legitimate server rejects the fake device $D_{k_F}$ because $D_{k_F}$ cannot generate the correct $r_j'$.

\subsection{Physical Attacks}\label{subsuc62}

In the proposed authentication protocol, the server is in such an insecure environment that an attacker can completely control the server. Therefore, an attacker can attack all the devices as well as all the servers in the authentication network.\par

1) The attacker attempts to launch intrusive probing attacks 
on legitimate servers to obtain internal configuration information 
for breaking the proposed authentication protocol.
However, probing attack on the servers is meaningless.
First, the server does not store any information related to the registered device, 
including the bit conversion function $Cover_k$ uniquely bound to the device $Device_k$, 
and the {LFSR-APUF}$_k$ uniquely embedded in $Device_k$.
Second, even if the server is broken during the authentication  
between server and $Device_k$, 
the attacker can only obtain the configuration information related to $Device_k$ 
and cannot obtain $Cover$ of other devices, 
which has no impact on other device nodes in the authentication network.
In addition, the attacker can only obtain $Cover_k$ through probing attacks, 
and information related to {LFSR-APUF}$_k$ cannot be obtained.
Therefore, the attacker cannot impersonate a server to launch spoofing attacks against $Device_k$.
Third, a legitimate server does not store any information related to mutual authentication internally, 
and essentially relies on encrypted information stored in databases such as $Enc_k$ and $Enc_{ki}$.
The server uses the keys $K_{s1}$ and $K_{s2}$, generated by \emph{WPUF}$_1$ and \emph{WPUF}$_2$, 
to decrypt the above encrypted message.
The attacker can control the server but cannot get $K_{s1}$ and $K_{s2}$ because
\emph{WPUF}$_1$ and \emph{WPUF}$_2$ are embedded in the server at the time of manufacturing, 
and any intrusive probing and tampering with the server can change the key values generated by $K_{s1}$ and $K_{s2}$ \cite{bib28}, 
making it difficult for the server to properly decrypt the encrypted information from the database.\par

2) The attacker attempts to launch an intrusive probing attack against legitimate $Device_k$. 
However, such a probing attack cannot succeed.
First, an attacker can obtain the $Cover_k$ circuit inside $Device_k$ by the probing attack, 
but the $Cover_k$ circuit cannot be generalized to other registered devices.
Second, for the underlying {LFSR-APUF}$_k$ circuit in $Device_k$, 
the {LFSR}$_k$ metal wires are attached to the delay path of \emph{APUF}$_k$.
Therefore, once an attacker detects the internal structure of {LFSR-APUF}$_k$ by unencapsulation, 
the deep submicron structure of \emph{APUF}$_k$ is destroyed, and the generated responses are inevitably changed.

\subsection{Modeling Attacks}\label{subsuc63}

The proposed authentication protocol can resist modeling attacks and brute force attacks against PUF.
Brute force attack can be classified as a modeling attack 
because its primary purpose is to obtain CRP information for modeling.
First, although the attacker has free access to the database, 
all CRP information inside the database is stored after encryption with $K_{s2}$.
Second, there is no explicit CRP transmission on the channel during the authentication protocol, 
and the CRP information about the PUF is basically covered by the proposed $Cover$ function.
Third, the PUF embedded in the device is not a standard APUF 
but an {LFSR-APUF} based on a challenge obfuscation mechanism, 
where {LFSR} and APUF are connected with each other.
The introduction of {LFSR} breaks the linear input-output mapping 
of the standard APUF circuit and greatly enhances the modeling attack resistance of the APUF.
In addition, the output of {LFSR-APUF} is not only related to the challenge $<c>_n$ but also to the $Base$.
Specifically, $Base$ value controls the shifting method of {LFSR}, 
i.e.\,, different $Base$ values correspond to different challenge obfuscation schemes, 
which changes the original static input-output mapping relationship of {LFSR-APUF}
to a dynamic input-output mapping relationship.

Fourth, although it seems impossible, if an attacker can directly obtain the CRPs of {LFSR-APUF}, 
he/she cannot model the PUF accurately to predict the unknown CRPs. 
For 64-stage {LFSR-APUF} with $Base$ value of 1, 
the prediction accuracy of LR, SVM, and ES is only slightly above 50\% 
(See Section \ref{subsec41} and Fig. \ref{fig8} for details).

\subsection{Comparison with Other PUF-Based Authentication Protocols}\label{subsuc64}

\begin{table*}
\centering
\caption{$Comparison$ $about$ $property$ $with$ $other$ $authentication$ $protocols$ $based$ $PUF$ }\label{tab3}
\begin{threeparttable}
\begin{tabular}{@{}lcccccccc@{}}
\hline
 & $SPUF$\cite{bib35} & $CSP$\cite{bib36} & $PUF-FSM$\cite{bib6} & $PUF-RAKE$\cite{bib37} & $MISR$\cite{bib32} &  $RSO$\cite{bib8} & $PUF-IPA$\cite{bib31} & $This\;Work$\\
\hline
P1\tnote{1}   & \checkmark  &  $X$  & $X$ & $X$ & $X$ & \checkmark & \checkmark  & $X$\\
P2\tnote{2} & $X$   &  $X$  & \checkmark & $X$ & $X$ & $X$ & \checkmark  & $X$\\
P3\tnote{3}   & $X$   &  $X$  & $X$ & $X$ & $X$ & $X$ & $X$  & $X$ \\
P4\tnote{4}   & $X$   & $X$  & \checkmark & $X$ & \checkmark & \checkmark & $X$  & $X$ \\
P5\tnote{5}   & $X$   &  \checkmark  & \checkmark & $X$ & $X$ & $X$ & $X$  & $X$ \\
P6\tnote{6}   & \checkmark   &  $X$  & $X$ & $X$ & \checkmark & \checkmark & $X$  & $X$ \\
P7\tnote{7}  & \checkmark   &   $X$  & \checkmark & \checkmark & $X$ & $X$ & \checkmark  & \checkmark\\
P8\tnote{8}   & \checkmark  &  \checkmark  & \checkmark & \checkmark & \checkmark & \checkmark & \checkmark  & \checkmark \\
P9\tnote{9}  & \checkmark   &  $X$  & $X$ & \checkmark & $X$ & $X$ & \checkmark  & \checkmark\\
P10\tnote{10}  & $X$   &  $X$  & $X$ & $X$ & $X$ & $X$ & $X$  & \checkmark\\
\hline
\end{tabular}
\begin{tablenotes}

\footnotesize
\item[1]{P1: Whether the device contains the NVM module in the authentication protocol.}
\item[2]{P2: Whether the device contains the hash logic in the authentication protocol.}
\item[3]{P3: Whether the device has the helper data leakage problem in the authentication protocol.}
\item[4]{P4: Whether the authentication protocol will be broken assuming the server is controlled by any attacker.}
\item[5]{P5: Whether the server contains a secure memory to store the CRPs of PUFs from all devices in the authentication network.}
\item[6]{P6: Whether the server contains a secure memory to store the soft models of PUFs from all devices in the authentication network.}
\item[7]{P7: Whether the authentication protocol realizes a mutual authentication.}
\item[8]{P8: Whether the proposed authentication protocol can resist modeling attacks effectively.}
\item[9]{P9: Whether the proposed authentication protocol can resist spoofing attacks including impersonating the server or impersonating the device.}
\item[10]{P10: Whether the proposed authentication protocol can resist physical attacks combined with spoofing attacks.}
\end{tablenotes}
\end{threeparttable}
\end{table*}

The characteristics of the proposed authentication protocol over other protocols are as follows: 

1) The device requires no NVM to store the key, which not only makes it suitable for low-cost platforms but also can resist probing attacks. Some devices based on PUF contain NVM to store some parameters used by the authentication protocol \cite{bib8}, \cite{bib31}, which greatly increases the resource overhead. Most importantly, NVM is vulnerable to physical attacks.
2) The device does not contain heavyweight hash logic but instead adopts the ultra-lightweight private bit conversion function $Cover$, which reduces the area overhead as well as the overall cost of the device significantly. 
3) The device does not use the fuzzy extractor as the error correction means, which reduces the hardware overhead and also ensures that the helper data of the device cannot be leaked, so the PUF is not threatened by the reliability-based side channel attack. 
4) Even if the server is completely controlled by the attacker, the attacker cannot obtain any data related to the PUF. In most of the previously reported PUF-based authentication protocols, data related to PUF is stored in the server \cite{bib8}, \cite{bib30}, \cite{bib33}, \cite{bib32}, so once the server is controlled by an attacker, he/she can obtain the PUF-related information and break the entire authentication network. 
5) The server does not store the CRPs from PUF of all the devices in the authentication network, but it stores the CRPs and other information in the database after encryption. This not only improves the security of authentication protocols but also eliminates the need for large and expensive on-server memory.
6) The server does not need to store the PUF soft model of all the devices in the authentication network, which reduces the server overhead. In addition, it is very difficult to establish a PUF soft model that is exactly the same as the PUF circuit of the device. The difference between the PUF soft model of the server and the PUF circuit of the device can increase the probability of authentication failure. 
7) Mutual authentication is fully realized, i.e.\,, device-to-server authentication and server-to-device authentication are both satisfied. Some authentication protocols only support server-to-device authentication \cite{bib33}, \cite{bib32}, \cite{bib8}, so they are vulnerable to spoofing attacks by fake servers, thereby destroying the entire authentication network. 
8) The proposed LFSR-APUF is embedded in the device and is deeply combined with the authentication protocol (See $Base$ in Section \ref{subsec32}) to realize the dynamic and time-variant obfuscation mechanism, which can effectively resist modeling attacks. Table III summarizes the discussed comparison.

\section{Conclusion}\label{sec7}

In this study, we proposed a lightweight authentication protocol against modeling attacks based on a novel {LFSR-APUF}. Firstly, an {LFSR-APUF} resistant to modeling attacks was established. The shift strategy of LFSR unit changed with the external challenges, realizing a dynamic and time-variant obfuscation by deep combination with the authentication protocol.The proposed {LFSR-APUF} could effectively prevent modeling attacks with a prediction rate of 51.79\%. Moreover, the introduced obfuscation scheme had no considerable impact on
the randomness, reliability, and uniqueness of the conventional APUF. Further, the corresponding strategy may be useful for implementing other APUF-based obfuscation schemes. Secondly, an ultra-lightweight PUF-based authentication protocol with a novel bit conversion function was proposed. In particular, the bit conversion $Cover$ function was private for every device in the authentication network, which further strengthened the security of the overall authentication network. The proposed authentication protocol ensured the security of the entire authentication network on the premise that the server could be completely controlled by any attacker. In addition, this authentication protocol took the proposed {LFSR-APUF} and private $Cover$ function as the encryption core, which facilitated an exclusive and unique encryption for every device in the authentication network. Overall, the proposed authentication protocol is a feasible and promising strategy for the security of IoT devices.

\bibliographystyle{IEEEtran}

\bibliography{ref}

\begin{IEEEbiography}[{\includegraphics[width=1in,height=1.25in,clip,keepaspectratio]{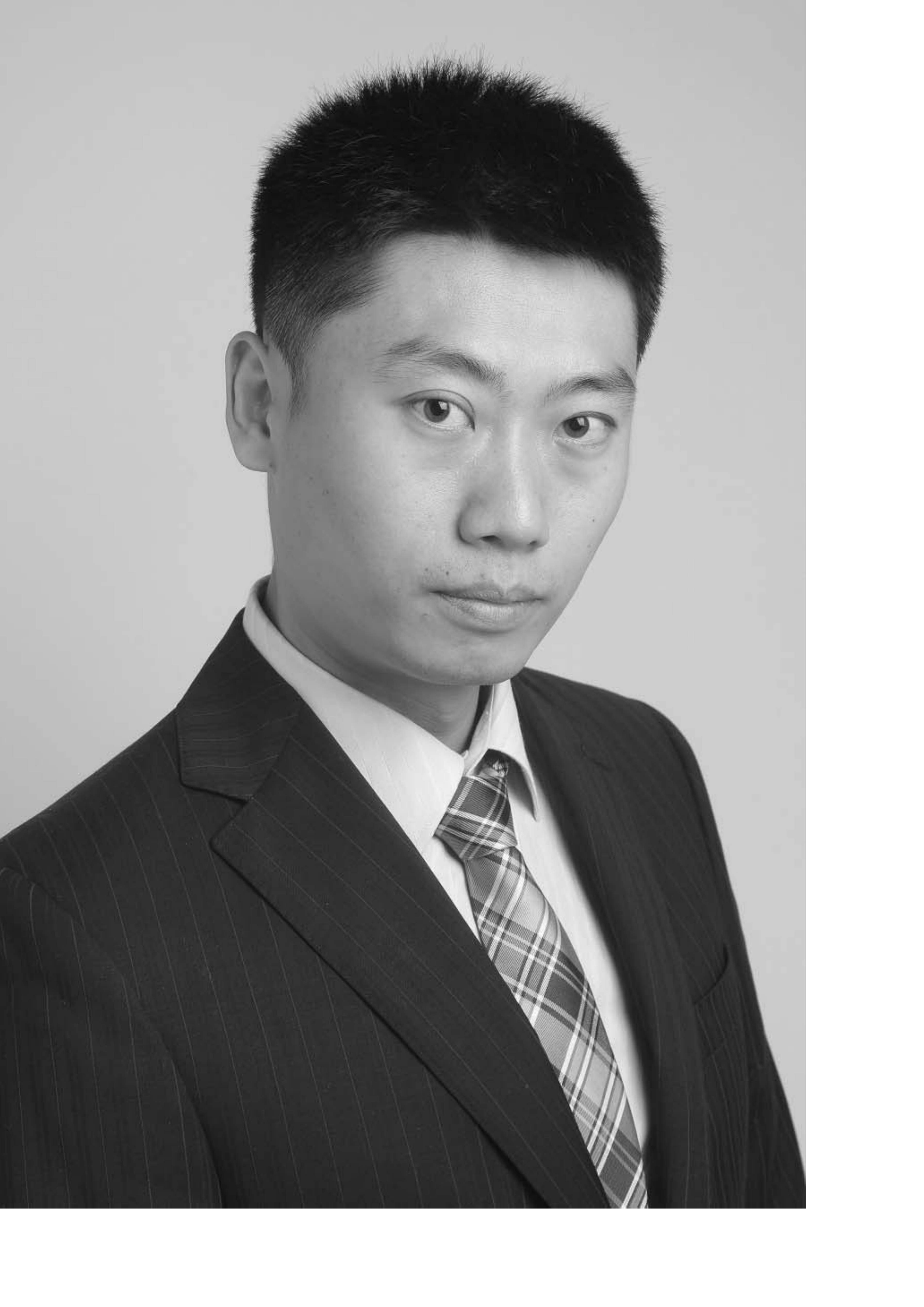}}]{Yao Wang}
(M’16) was born in Henan, China, in 1983. He received his M.S. degree from Zhengzhou University, Zhengzhou, China, in 2009, and the Ph.D. degree from the University of Electronic Science and Technology of China (UESTC), Chengdu, China, in 2013.
From 2014 to 2017, he was a Lecturer with UESTC, then he joined Zhengzhou University. Since 2018, he has been an Associate Professor at the School of Information Engineering, Zhengzhou University. His research interests include low-power mixed-signal integrated circuits and hardware security for IoT applications . 
\end{IEEEbiography}

\begin{IEEEbiography}[{\includegraphics[width=1in,height=1.25in,clip,keepaspectratio]{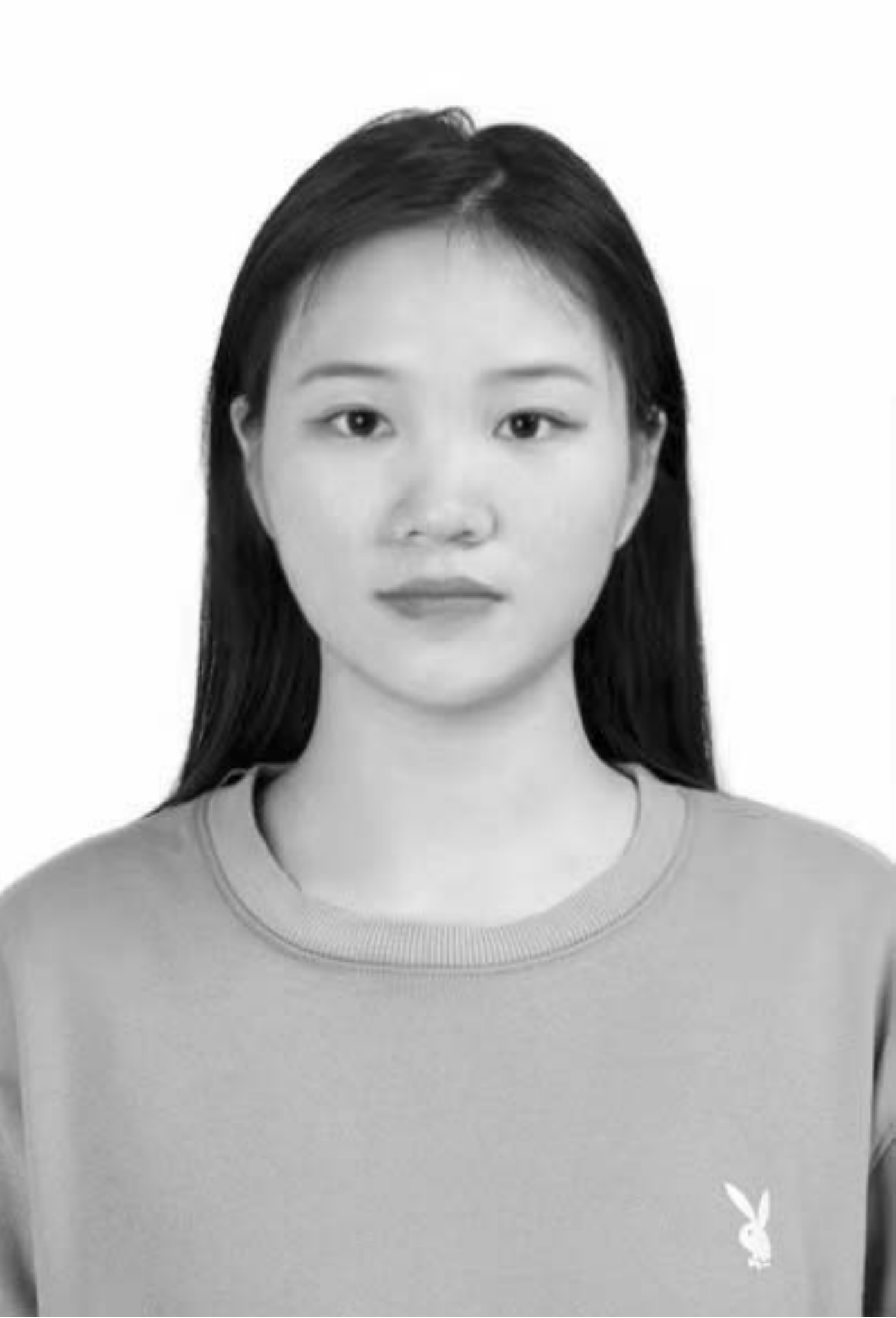}}]{Xue Mei}
 was born in 1998 in Henan, China. She received a Bachelor of Science degree. She graduated from Xinxiang of Henan Normal University in China in 2020. She is currently pursuing the M.S. degree at Zhengzhou University in China. Her current research interests include hardware security and digital integrated circuits.
\end{IEEEbiography}

\begin{IEEEbiography}[{\includegraphics[width=1in,height=1.25in,clip,keepaspectratio]{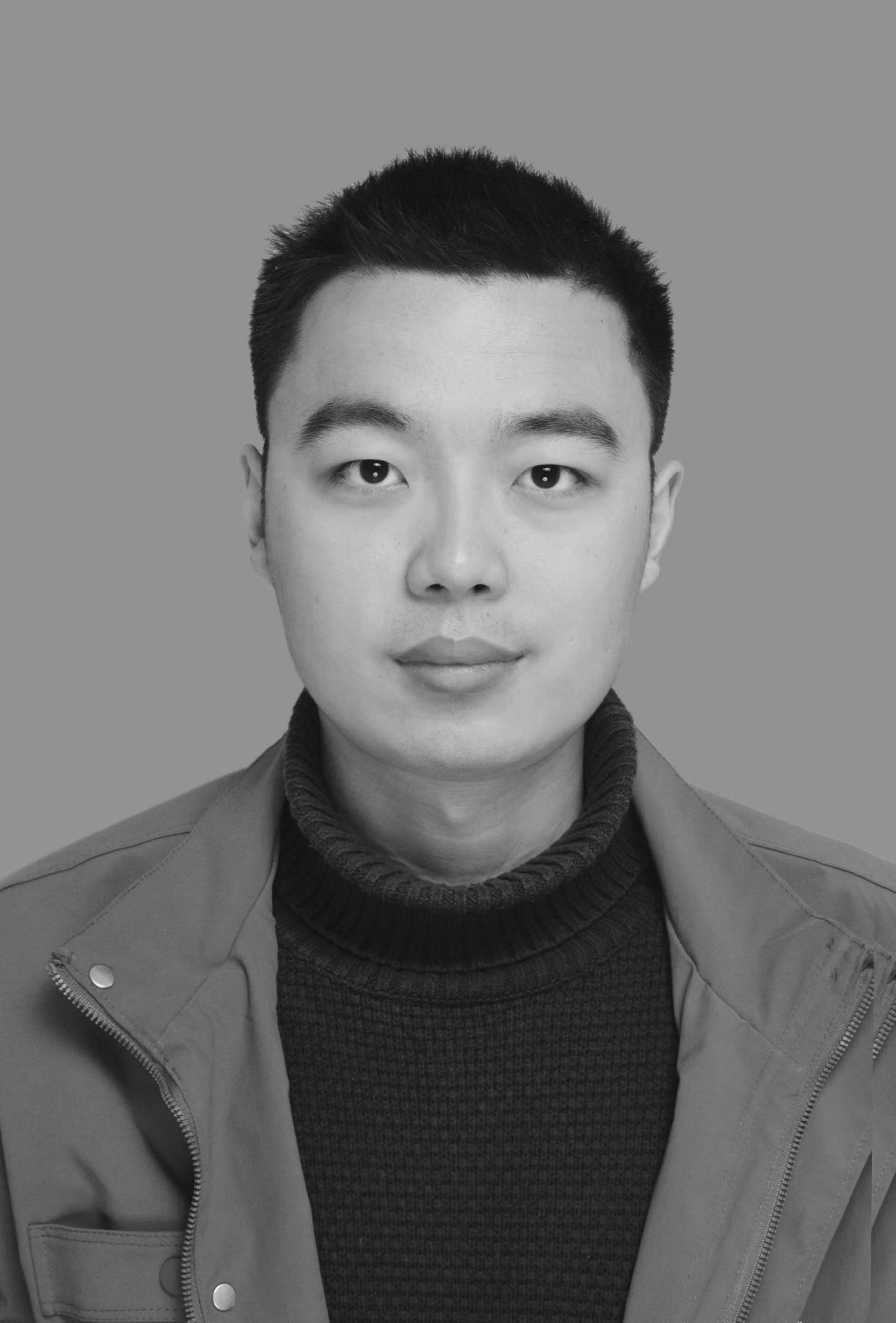}}]{Zhengtai Chang}
was born in Henan, China, in 1997. He received his B.S. degree from Henan Agricultural University in 2019, and the M.S. degree from Zhengzhou University in 2022, respectively. He is currently an FPGA engineer at XJ Group Corporation. His research interests include hardware security and machine learning.
\end{IEEEbiography}

\begin{IEEEbiography}[{\includegraphics[width=1in,height=1.25in,clip,keepaspectratio]{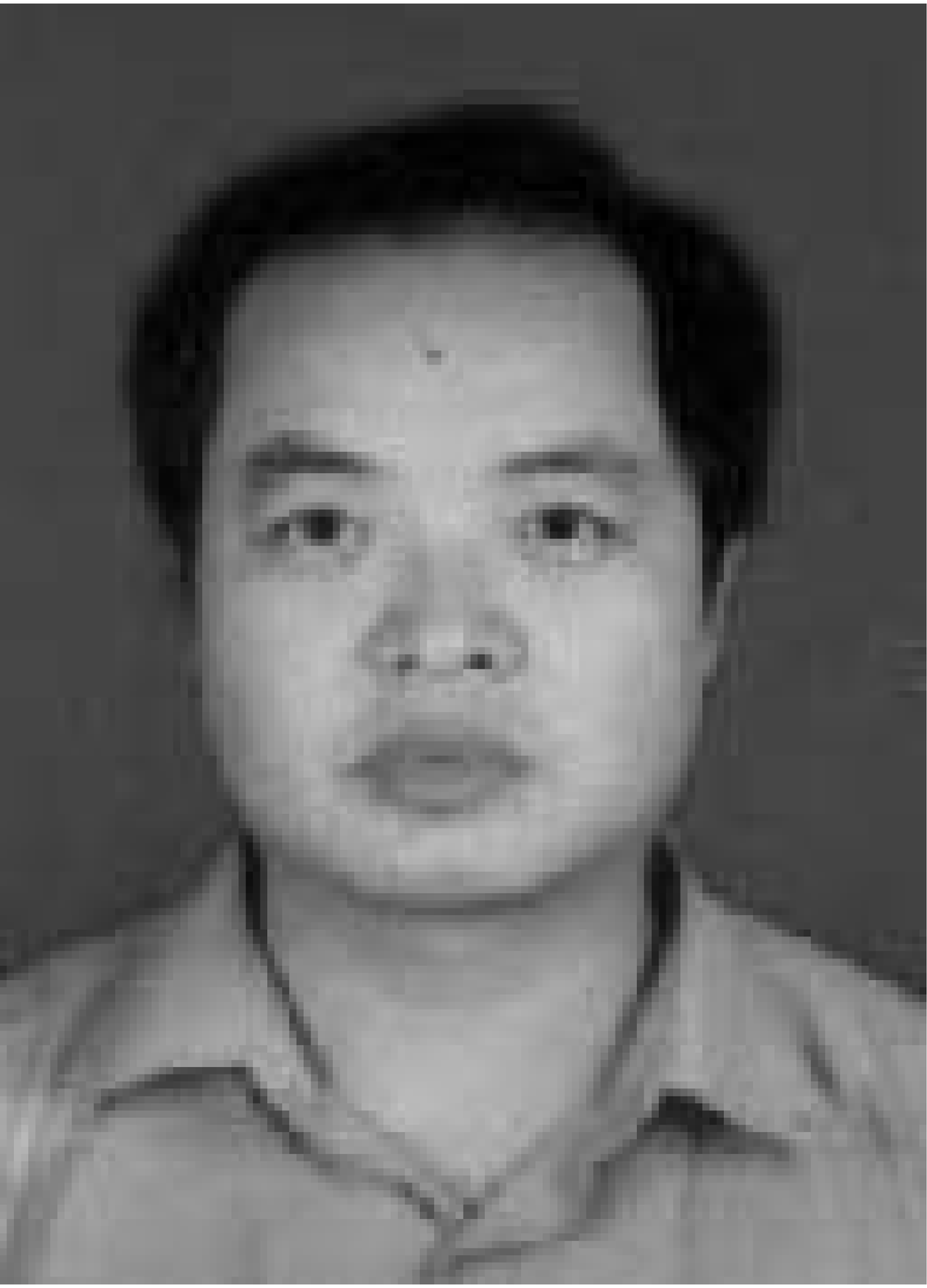}}]{Wenbing Fan}
received the Ph.D. degree from East China University of Science and Technology, Shanghai, China, in 2003. He is currently a Professor at the School of Information Engineering, Zhengzhou University, Zhengzhou, China. His current research interests include mixed signal circuits and embedded systems.
\end{IEEEbiography}

\begin{IEEEbiography}[{\includegraphics[width=1in,height=1.25in,clip,keepaspectratio]{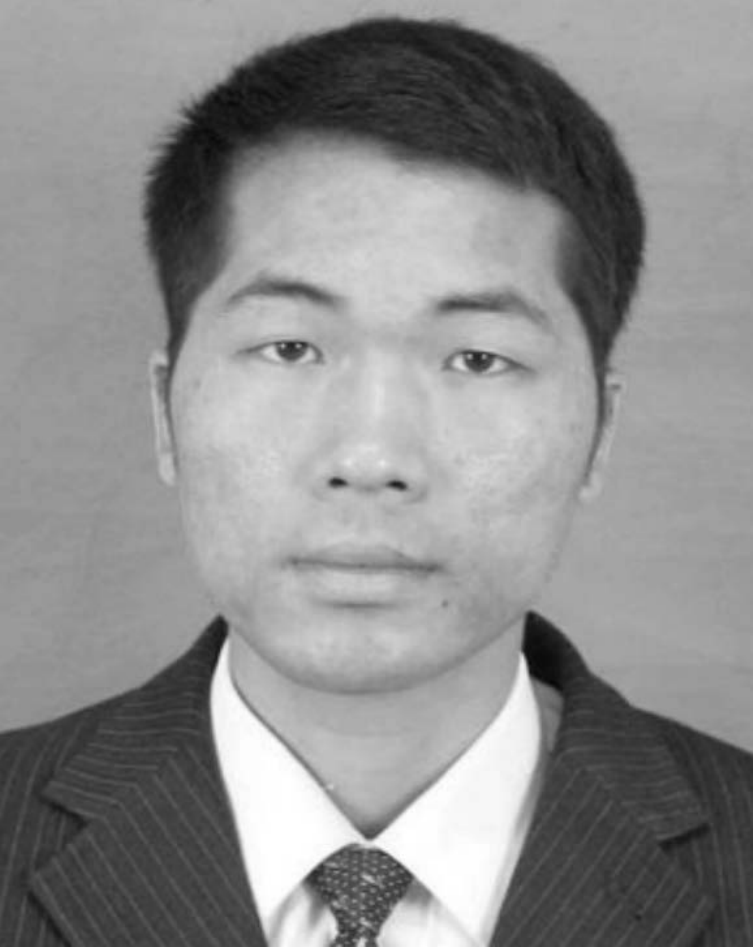}}]{Benqing Guo}
received the M.Sc. and Ph.D. degrees in electrical engineering from UESTC, Chengdu, China, in 2005 and 2011, respectively.
He is currently a Researcher with UESTC. His current research interests include RF/analog integrated circuits and techniques.
Dr. Guo served as a Technical Reviewer for over 15 international journals and conferences, including IEEE Trans. MTT, IEEE Trans. VLSI, and IEEE ISCAS.  
\end{IEEEbiography}

\begin{IEEEbiography}[{\includegraphics[width=1in,height=1.25in,clip,keepaspectratio]{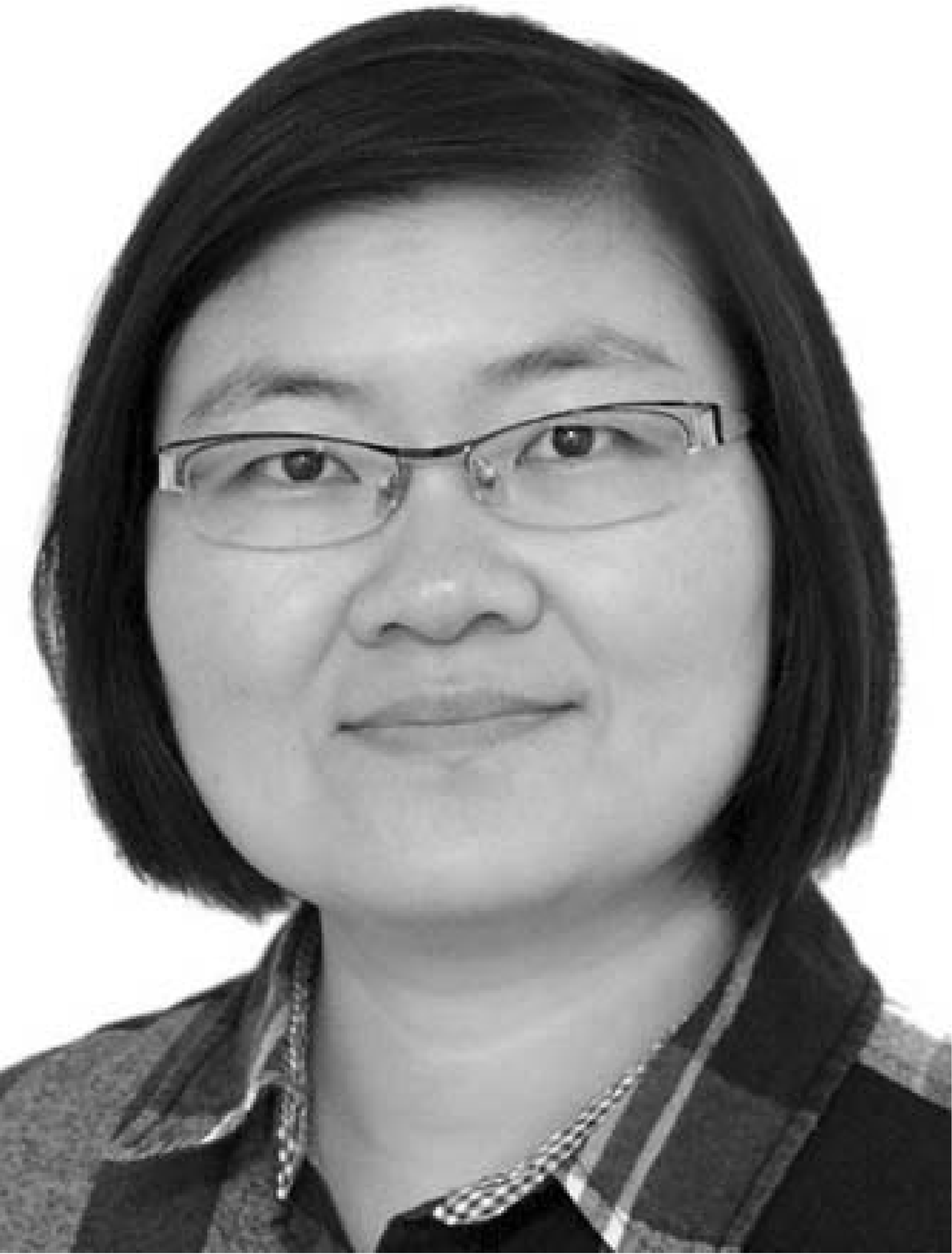}}]{Zhi Quan}
received the B.S. in electronics engineering from Hunan University, China, in 2001 and the Ph.D. degree in electronics engineering from University of York, the United Kingdom, in 2009. She then joined the Department of Electrical Engineering of the Federal University of Juiz de Fora, Brazil as a postdoctoral researcher. Currently, she is a professor at the School of Information Engineering, Zhengzhou University. Her research interests include high speed digital circuits, FPGA techniques and power line communication.
\end{IEEEbiography}



\end{document}